\documentclass[10pt]{IEEEtran}
\usepackage{graphicx}
\usepackage{tikz}
\usepackage{comment}
\usetikzlibrary{positioning,spy}
\usepackage{amsmath,amsfonts,amssymb,amsthm}
\usepackage{booktabs}
\usepackage{algorithm,algorithmic}
\usepackage{bm}
\usepackage{url}
\usepackage[caption=false,font=footnotesize]{subfig}

\DeclareMathOperator*{\argmin}{arg\,min}
\DeclareMathOperator*{\sgn}{sgn}
\DeclareMathOperator*{\mse}{MSE}

\def\bx{\ensuremath{{\bf x}}}
\def\bm{\ensuremath{{\bf m}}}
\def\bg{\ensuremath{{\bf g}}}
\def\by{\ensuremath{{\bf y}}}
\def\bz{\ensuremath{{\bf z}}}
\def\bw{\ensuremath{{\bf w}}}

\def\bk{\ensuremath{{\bf k}}}
\def\bn{\ensuremath{{\bf n}}}
\def\bc{\ensuremath{{\bf c}}}
\def\bb{\ensuremath{{\bf b}}}
\def\bd{\ensuremath{{\bf d}}}
\def\bP{\ensuremath{{\bf P}}}

\def\bR{\ensuremath{{\bf R}}}
\def\bF{\ensuremath{{\bf F}}}
\def\bI{\ensuremath{{\bf I}}}

\def\bone{\ensuremath{{\bf 1}}}
\def\diag{\ensuremath{{\mathrm{diag}}}}

\def\cT{\ensuremath{{\mathcal T}}}
\def\cF{\ensuremath{{\mathcal F}}}
\def\cI{\ensuremath{{\mathcal I}}}
\def\cS{\ensuremath{{\mathcal S}}}
\def\cL{\ensuremath{{\mathcal L}}}

\title{Deep Algorithm Unrolling for \\ Blind Image Deblurring}

\author{Yuelong~Li,~\IEEEmembership{Student~Member,~IEEE,}
	Mohammad~Tofighi,~\IEEEmembership{Student~Member,~IEEE,}
	Junyi~Geng,~\IEEEmembership{Student~Member,~IEEE,}
	Vishal~Monga,~\IEEEmembership{Senior~Member,~IEEE,}
	and Yonina C. Eldar,~\IEEEmembership{Fellow,~IEEE}

	\thanks{Y. Li, M. Tofighi, and V. Monga are with Department of Electrical Engineering, The Pennsylvania State University, University Park,
		PA, 16802 USA, Emails: yul200@psu.edu, tofighi@psu.edu, vmonga@engr.psu.edu
	}
	\thanks{Y. C. Eldar is with Department of Electrical Engineering, Technion, Israel Institute of Technology, Haifa, Israel, Email: yonina@ee.technion.ac.il}

}

\begin{document}

\maketitle

\begin{abstract}
Blind image deblurring remains a topic of enduring interest. Learning based approaches, especially those that employ neural networks have emerged to complement traditional model based methods and in many cases achieve vastly enhanced performance. That said, neural network approaches are
generally empirically designed and the underlying structures are
difficult to interpret. In recent years, a promising technique called
algorithm unrolling has been developed that has helped connect iterative algorithms such as those for sparse coding to neural network architectures.
However, such connections have not been made yet for blind image deblurring. In this paper, we propose a neural network architecture
based on this idea. We first present an iterative algorithm that may
be considered as a generalization of the traditional total-variation
regularization method in the gradient domain. We then unroll
the algorithm to construct a neural network for image deblurring which we refer to as Deep Unrolling for Blind Deblurring (DUBLID). Key algorithm parameters are learned with the help of training images. Our proposed deep network DUBLID
achieves significant practical performance gains while
enjoying interpretability at the same time. Extensive experimental
results show that DUBLID outperforms many state-of-the-art
methods and in addition is computationally faster.\vspace{-4pt}
\end{abstract}

\section{Introduction}\label{sec:introduction}
\IEEEPARstart{B}{lind} image deblurring refers to the process of recovering a
sharp image from its blurred observation without explicitly knowing the blur function. In real world imaging, images
frequently suffer from degraded quality as a consequence of blurring artifacts,
which blind deblurring algorithms are designed to remove such artifacts. These
artifacts may come from different sources, such as atmospheric turbulence,
diffraction, optical defocusing, camera shaking,
and more~\cite{kundur_blind_1996}.  In the computational imaging literature,
motion deblurring is an important topic because camera shakes are
common during the photography procedure. In recent years, this topic has
attracted growing attention thanks to the popularity of smartphone cameras.  On
such platforms, the motion deblurring algorithm plays an especially crucial
role because effective hardware solutions such as professional camera stabilizers are difficult to
deploy due to space restrictions.

In this work we focus on motion deblurring in particular because of its
practical importance. However, our development does not make assumptions on
blur type and hence may be extended to cases other than motion blur.  Motion
blurs occur as a consequence of relative movements between the camera and the
imaged scene during exposure. Assuming the scene is planar and the camera
motion is translational, the image degradation process may be modelled as
a discrete convolution~\cite{kundur_blind_1996}: \begin{equation}
\by=\bk\ast\bx+\bn,\label{eq:blur_model} \end{equation} where $\by$ is the
observed blurry image, $\bx$ is the latent sharp image, $\bk$ is the unknown point spread
function (blur kernel), and $\bn$ is random noise which is often modelled
as Gaussian. Blind motion deblurring corresponds to estimating both $\bk$
and $\bx$ given $\by$; this estimation problem is also commonly called blind
deconvolution.

\noindent\textbf{Related Work:} The majority of existing blind motion deblurring methods are based on
iterative optimization. Early works can be traced back to
several decades
ago~\cite{richardson_bayesian-based_1972,shepp_maximum_1982,ayers_iterative_1988,kundur_blind_1996,chan_total_1998}.
These methods are only effective when the blur kernel is relatively
small. In the last decade, significant breakthroughs have been made both
practically and conceptually. As both the image and the kernel need
to be estimated, there are infinitely many pairs of solutions forming the
same blurred observation rendering blind deconvolution an ill-posed
problem. A popular remedy is to add regularizations so that many blind deblurring
algorithms essentially reduce to solving regularized inverse problems.
A vast majority of these techniques hinge on sparsity-inducing regularizers,
either in the gradient
domain~\cite{joshi_psf_2008,shan_high-quality_2008,cho_fast_2009,xu_two-phase_2010,krishnan_blind_2011,xu_unnatural_2013,sun_edge-based_2013,pan_$l_0$_2017}
or more general sparsifying transformation
domains~\cite{jian-feng_cai_framelet-based_2012,xiang_image_2015,pan_deblurring_2018,tofighi2018blind}.
Variants of such methods may arise indirectly from a statistical
estimation perspective, e.g.~\cite{fergus_removing_2006,levin_efficient_2011,babacan_bayesian_2012,wipf_revisiting_2014}.

From a conceptual perspective, Levin {\it et
al.}~\cite{levin_understanding_2011} study the limitations and
remedies of the commonly employed Maximum a Posterior (MAP) approach,
while Perrone {\it et al.}~\cite{perrone_clearer_2016} extend their
study with a particular focus on Total-Variation (TV) regularization.
Despite some performance improvements achieved along their
developments, the iterative optimization approaches generally suffer from several
limitations. First, their performance depends heavily on
appropriate selection of parameter values. Second, handcrafted
regularizers play an essential role, and designing versatile regularizers that generalize well to a variety of real datasets can be a
challenging task.  Finally, hundreds and thousands of iterations are
often required to reach an acceptable performance level and thus
these approaches can be slow in practice.

Complementary to the aforementioned approaches, learning based methods for determining a non-linear mapping that deblurs the image while adapting parameter choices to an underlying training image set have been developed. Principally important in this class are techniques that employ deep neural networks. The history of leveraging neural networks for blind deblurring actually dates back to
the last century~\cite{steriti_blind_1994}. In the past few years, there has been a
growing trend in applying neural networks to various imaging
problems~\cite{lucas_using_2018}, and blind motion deblurring has followed that trend. Xu {\it et al.}~\cite{xu_deep_2014} use large convolution kernels
with carefully chosen initializations in a Convolutional Neural Network (CNN); Yan {\it
et al.}~\cite{yan_blind_2016} concatenate a classification network with a
regression network to deblur images without prior information about the blur
kernel type. Chakrabarti {\it et al.}~\cite{chakrabarti_neural_2016} work in
the frequency domain and employ a neural network to predict Fourier transform
coefficients of image patches; Xu {\it et al.}~\cite{xu_motion_2018} employ a
CNN for edge enhancement prior to kernel and image estimation. These works
often outperform iterative optimization algorithms especially for linear motion kernels;
however, the structures of the
networks are often empirically determined and their actual functionality is hard to interpret.\vspace{-2pt}

In the seminal work of Gregor {\it et
al.}~\cite{gregor_learning_2010}, a novel technique called algorithm
unrolling was proposed. Despite its focus on approximating sparse
coding algorithms, it provides a principled framework for expressing
traditional iterative algorithms as neural networks, and offers
promise in developing interpretable network architectures.
Specifically, each iteration step may be represented as one layer of
the network, and concatenating these layers form a deep neural
network. Passing through the network is equivalent to executing the
iterative algorithm a finite number of times. The network may be
trained using back-propagation~\cite{lecun_gradient-based_1998}, and
the trained network can be naturally interpreted as a parameter
optimized algorithm. An additional benefit is that prior knowledge
about the conventional algorithms may be transferred. There has been limited recent exploration of neural network architectures by unrolling iterative
algorithms for problems such as super-resolution and clutter/noise suppression ~\cite{wang_deep_2015,jin_deep_2017,chen_trainable_2017,solomon_deep_2018}.
In blind deblurring, Schuler {\it et
al.}~\cite{schuler_learning_2016} employ neural networks as feature
extraction modules towards a trainable deblurring
system.  However, the network portions are still empirical. \vspace{-2pt}

Other aspects of deblurring have been investigated such as 
spatially varying blurs~\cite{tai_richardson-lucy_2011,whyte_non-uniform_2012}, including some recent neural network
approaches~\cite{sun_learning_2015,nah_deep_2017,nimisha_blur-invariant_2017,su_deep_2017}.
Other algorithms benefit from device
measurements~\cite{raskar_coded_2006,cho_motion_2010,joshi_image_2010} or
leverage multiple images~\cite{cai_blind_2009,sroubek_robust_2012}.\vspace{-1pt}

\noindent\textbf{Motivations and Contributions:} Conventional iterative algorithms have the merits of interpretability, but
acceptable performance levels demand much longer execution time compared to modern neural network approaches. Despite previous efforts, the link between both
categories remains largely unexplored for the problem of blind deblurring, and a
method that simultaneously enjoys the benefits of both is lacking. In this regard, we make the following contributions:\footnote{A preliminary 4 page version of this work has been submitted to IEEE ICASSP 2019 \cite{li_icassp19}. This paper  involves substantially more analytical development in the form of: a.) the unrolling mechanism and associated optimization problem for learning parameters, b.) derivation of custom back-propagation rules, c.) handling of color images, and d.) demonstration of computational benefits. Experimentally, we have added a new dataset and several new state of the art methods and scenarios in our comparisons. Finally, ablation studies have been included to better explain the proposed DUBLID and its merits.}
\begin{itemize}
	\item \textbf{Deep Unrolling for BLind Deblurring (DUBLID):} We propose an interpretable neural network structure called DUBLID. We first present an iterative algorithm that may
	be considered a generalization of the traditional total-variation
	regularization method in the gradient domain, and subsequently unroll
	the algorithm to construct a neural network. Key algorithm parameters are learned with the help of training images using backpropagation, for which we derive analytically simple forms that are amenable to fast implementation.

	\item \textbf{Performance and Computational Benefits:} Through extensive experimental validation over {\em three} benchmark datasets, we verify the superior
		performance of the proposed DUBLID, both over conventional iterative
		algorithms and more recent neural network approaches. Both traditional linear and more recently developed non-linear kernels are used in our experiments. Besides quality gains, we show that DUBLID is computationally simpler. In particular, the carefully designed interpretable layers enables DUBLID to learn with far fewer parameters than state of the art deep learning approaches -- hence leading to much faster inference time.
			
	\item \textbf{Reproducibility:} To ensure reproducibility, we share our code and datasets that are used to generate all our experimental results freely
		online.
\end{itemize}

The rest of the paper is organized as follows.  Generalized gradient domain
deblurring is reviewed in Section~\ref{sec:formulation}. We identify the roles
of (gradient/feature extraction) filters and other key parameters, which are
usually assumed {\em fixed}.  Based on a half-quadratic optimization procedure
to solve the aforementioned gradient domain deblurring, we develop a new
unrolling method that realizes the iterative optimization as a neural network
in Section~\ref{sec:implementation}. In particular, we show that the various
linear and non-linear operators in the optimization can be cascaded to generate
an interpretable deep network, such that the number of layers in the network
corresponds to the number of iterations. The fixed filters and parameters are
now {\em learnable\/} and a custom back-propagation procedure is proposed to
optimize them based on training images.  Experimental results that provide
insights into DUBLID as well as comparisons with state of the art methods
are reported in Section~\ref{sec:experiments}. Section~\ref{sec:conclusion}
concludes the paper.\vspace{-3pt}

\section{Generalized Blind Deblurring via Iterative Minimization: A Filtered Domain Regularization Perspective}\label{sec:formulation}

\subsection{Blind Deblurring in the Filtered Domain}\label{subsec:formulation}
A common practice for blind motion deblurring is to estimate the
kernel in the image gradient
domain~\cite{fergus_removing_2006,shan_high-quality_2008,cho_fast_2009,xu_two-phase_2010,levin_efficient_2011,krishnan_blind_2011,perrone_clearer_2016}.
Because the gradient operator $\nabla$ commutes with convolution, 
taking derivatives on both side of~\eqref{eq:blur_model} gives
\begin{equation}
	\nabla\by=\bk\ast\nabla\bx+\bn^\prime,\label{eq:gradient_domain}
\end{equation}
where $\bn^\prime=\nabla\bn$ is Gaussian noise. Formulation in
the gradient domain, as opposed to the pixel domain, has several
desirable strengths: first, the kernel generally serves as a low-pass
filter, and low-frequency components of the image are barely
informative about the kernel.  Intuitively, the kernel may be
inferred along edges rather than homogeneous regions in the image.
Therefore, a gradient domain approach can lead to improved
performance in practice~\cite{levin_efficient_2011} as the gradient operator effectively filtered out the uninformative
regions. Additionally, from a
computational perspective, gradient domain formulations help in better
conditioning of the linear system resulting in more reliable estimation~\cite{cho_fast_2009}.

The model~\eqref{eq:gradient_domain} alone, however, is insufficient
for recovering both the image and the kernel; thus regularizers on both
are needed. The gradients of natural images are generally
sparse, i.e., most of their values are of small
magnitude~\cite{fergus_removing_2006,levin_understanding_2011}. This
fact motivates the developments of various sparsity-inducing
regularizations on $\nabla\bx$. Among them one of particular interest
is the $\ell_1$-norm (often called TV) thanks to its
convexity~\cite{chan_total_1998,perrone_clearer_2016}. To
regularize the kernel, it is common practice to assume the kernel
coefficients are non-negative and of unit sum. Consolidating these
facts, blind motion deblurring may be carried out by solving the
following optimization problem~\cite{perrone_clearer_2016}:\vspace{-7pt}
\begin{align}
	\min_{\bk,\bg_1,\bg_2}&\frac{1}{2}\left(\left\|D_x\by-\bk\ast\bg_1\right\|_2^2+\left\|D_y\by-\bk\ast\bg_2\right\|_2^2\right)\nonumber\\
						  &+\lambda_1\|\bg_1\|_1+\lambda_2\|\bg_2\|_1+\frac{\epsilon}{2}\|\bk\|_2^2,\nonumber\\
	\text{subject to }&\|\bk\|_1=1,\quad\bk\geq 0,\label{eq:gradient_optimization}\vspace{-7pt}
\end{align} 
where $D_x\by,D_y\by$ are the partial derivates of $\by$ in
horizontal and vertical directions respectively. The notation $\|\cdot\|_p$ denotes the $\ell_p$
vector norm, while $\lambda_1,\lambda_2,\varepsilon$ are positive
constant parameters to balance the contributions of each term. The
$\geq$ sign is to be interpreted elementwise.
The solutions $\bg_1$ and $\bg_2$ of~\eqref{eq:gradient_optimization} are estimates of
the gradients of the sharp image $\bx$, i.e., we may expect
$\bg_1\approx D_x\bx$ and $\bg_2\approx D_y\bx$.

In practice, numerical gradients of images are usually computed using
discrete filters, such as the Prewitt and Sobel filters. From this
viewpoint, $D_x\by$ and $D_y\by$ may be viewed as filtering $\by$
through two derivative filters of orthogonal directions~\cite{gonzalez2002digital}. Therefore, a
straightforward generalization of~\eqref{eq:gradient_optimization} is
to use more than two filters, i.e., pass $\by$ through a filter bank.
This generalization increases the flexibility
of~\eqref{eq:gradient_optimization}, and appropriate choice of the
filters can significantly boost performance. In particular, by
steering the filters towards more directions other than horizontal
and vertical, local features (such as lines, edges and textures) of
different orientations are more effectively
captured~\cite{freeman_design_1991,starck_curvelet_2002,unser_steerable_2011}.
Moreover, the filter can adapt its shapes to enhance the
representation
sparsity~\cite{mailhe_shift-invariant_2008,barthelemy_shift_2012}, a desirable property to pursue.

Suppose we have determined a desired collection of $C$ filters ${\{\mathbf{f}_i\}}_{i=1}^C$. By commutativity of convolutions, we have\vspace{-7pt}
\begin{equation}
	\mathbf{f}_i\ast\by=\mathbf{f}_i\ast\bk\ast\bx+\bn_i^\prime=\bk\ast(\mathbf{f}_i\ast\bx)+\bn_i^\prime,\quad i=1,2,\dots,C,\vspace{-7pt}\label{eq:observation_model}
\end{equation}
where the filtered noises $\bn_i^\prime=\mathbf{f}_i\ast\bn$ are
still Gaussian. To encourage sparsity of the filtered image, we formulate the
optimization problem (which may similarly be regarded as a generalization of~\cite{perrone_clearer_2016})\vspace{-5pt}
\begin{align}
	\min_{\bk,{\{\bg_i\}}_{i=1}^C}&\sum_{i=1}^C\left(\frac{1}{2}\left\|\mathbf{f}_i\ast\by-\bk\ast\bg_i\right\|_2^2+\lambda_i\|\bg_i\|_1\right)+\frac{\epsilon}{2}{\|\bk\|}_2^2,\nonumber\\
	\text{subject to }&\|\bk\|_1=1,\quad\bk\geq 0.\label{eq:objective}\vspace{-5pt}
\end{align}

\subsection{Efficient Minimization via Half-quadratic Splitting}\label{subsec:optimization}
Problem~\eqref{eq:objective} is non-smooth so that traditional
gradient-based optimization algorithms cannot be considered.
Moreover, to facilitate the subsequent unrolling procedure, the
algorithm needs to be simple (to simplify the network structure) and
converge quickly (to reduce the number of layers required). Based on
these concerns, we adopt the half-quadratic splitting
algorithm~\cite{wang_new_2008}. This algorithm is simple but
effective, and has been successfully employed in many previous
deblurring
techniques~\cite{schmidt_discriminative_2013,pan_$l_0$_2017,pan_deblurring_2018}.

The basic idea is to perform variable-splitting and then alternating
minimization on the penalty function. To this end, we first cast~\eqref{eq:objective} into the following approximation model:
\begin{align}
	\min_{\bk,{\{\bg_i,\bz_i\}}_{i=1}^C}&\sum_{i=1}^C\left(\frac{1}{2}{\left\|\mathbf{f}_i\ast\by-\bk\ast\bg_i\right\|}_2^2\right.\nonumber\\
	+&\left.\lambda_i{\|\bz_i\|}_1+\frac{1}{2\zeta_i}{\|\bg_i-\bz_i\|}_2^2\right)+\frac{\epsilon}{2}{\|\bk\|}_2^2,\nonumber\\
	\text{subject to }&\|\bk\|_1=1,\quad\bk\geq 0,\label{eq:penalty}
\end{align}
by introducing auxiliary variables ${\{\bz_i\}}_{i=1}^C$, where $\zeta_i,i=1,\dots,C$ are regularization parameters. It is well known that
as $\zeta_i\rightarrow 0$ the sequence of solutions to~\eqref{eq:penalty}
converges to that of~\eqref{eq:objective}~\cite{bertsekas2014constrained}.  In a similar manner to~\cite{pan_$l_0$_2017}, we
then alternately minimize over ${\{\bg_i\}}_{i=1}^C,{\{\bz_i\}}_{i=1}^C$ and
$\bk$ and iterate until convergence~\footnote{In the non-blind deconvolution literature, a formal convergence proof has been shown in~\cite{wang_new_2008}, while for blind deconvolution, empirical convergence has been frequently observed as shown in~\cite{xu_unnatural_2013,pan_$l_0$_2017}, etc.}. Specifically, at the $l$-th
iteration, we execute the following minimizations sequentially:
\begin{align}
	\bg_i^{l+1}\gets&\argmin_{\bg_i}\frac{1}{2}\left\|\mathbf{f}_i\ast\by-\bk^l\ast\bg_i\right\|_2^2+\frac{1}{2\zeta_i}\left\|\bg_i-\bz_i^l\right\|_2^2,\quad\forall i,\nonumber\\
	\bz_i^{l+1}\gets&\argmin_{\bz_i}\frac{1}{2\zeta_i}\left\|\bg^{l+1}_i-\bz_i\right\|_2^2+\lambda_i\|\bz_i\|_1,\quad\forall i,\nonumber\\
	\bk^{l+1}\gets&\argmin_\bk\sum_{i=1}^C\frac{1}{2}\left\|\mathbf{f}_i\ast\by-\bk\ast\bg_i^{l+1}\right\|_2^2+\frac{\epsilon}{2}\left\|\bk\right\|^2_2,\nonumber\\
							&\text{subject to }\|\bk\|_1=1,\quad\bk\geq 0.\label{eq:fixed}
\end{align}

For notational brevity, we will consistently use $i$ to index the filters and $l$ to index the layers (iteration) henceforth. The notations ${\{\}}_i$ and ${\{\}}_l$ collects every filter and layer components, respectively.
As it is, problem~\eqref{eq:objective} is non-convex over the joint variables
$\bk$ and ${\{\bg_i\}}_i$ and proper initialization is crucial to get good
	solutions. However, it is difficult to find appropriate
	initializations that perform well under various practical scenarios. An
alternative strategy that has been commonly employed
is to use different parameters per
iteration~\cite{wang_new_2008,xu_unnatural_2013,perrone_clearer_2016,pan_$l_0$_2017}.
For example, $\lambda_i$'s are typically chosen as a large value from the
beginning, and then gradually decreased towards a small constant.
In~\cite{wang_new_2008} the values of $\zeta_i$'s decrease as the algorithm
proceeds for faster convergence. In numerical analysis and optimization, this
strategy is called the continuation method and its
effectiveness is known for solving non-convex 
problems~\cite{blake_visual_1987}.  By adopting this strategy, we choose
different parameters ${\{\zeta_i^l,\lambda_i^l\}}_{i,l}$ across the iterations.
We take this idea one step further by optimizing the filters across iterations as well, i.e. we design filters
${\{\mathbf{f}^l_i\}}_{i,l}$.  Consequently, the alternating minimization
scheme in~\eqref{eq:fixed} becomes:
\begin{align}
	\bg_i^{l+1}\gets&\argmin_{\bg_i}\frac{1}{2}\left\|\mathbf{f}^l_i\ast\by-\bk^l\ast\bg_i\right\|_2^2+\frac{1}{2\zeta^l_i}\left\|\bg_i-\bz_i^l\right\|_2^2,\quad\forall i,\label{eq:gstep}\\
	\bz_i^{l+1}\gets&\argmin_{\bz_i}\frac{1}{2\zeta^l_i}\left\|\bg^{l+1}_i-\bz_i\right\|_2^2+\lambda^l_i\|\bz_i\|_1,\quad\forall i,\label{eq:zstep}\\
	\bk^{l+1}\gets&\argmin_\bk\sum_{i=1}^C\frac{1}{2}\left\|\mathbf{f}^l_i\ast\by-\bk\ast\bg_i^{l+1}\right\|_2^2+\frac{\epsilon}{2}\left\|\bk\right\|^2_2,\label{eq:kstep}\\
					   &\text{subject to }\|\bk\|_1=1,\quad\bk\geq 0.\nonumber
\end{align}

\noindent We summarize the complete algorithm in Algorithm~\ref{alg:half_quadratic},
where $\delta$ in Step~\ref{step:init} is
the impulse function. Problem~\eqref{eq:gstep} can be efficiently solved by making use of the Discrete
Fourier Transform (DFT) and its solution is given in Step~\ref{step:gupdate} of Algorithm~\ref{alg:half_quadratic}, where $\cdot^\ast$ is the complex conjugation and $\odot$
is the Hadamard (elementwise) product operator. The
operations are to be interpreted elementwise when acting on matrices
and vectors. We let $\widehat{\cdot}$
denote the DFT and $\cF^{-1}$ be the inverse DFT.\@ The closed-form solution to problem~\eqref{eq:zstep}
is well known and can be found in Step~\ref{step:threshold}, where, $\cS_\lambda(\cdot)$ is the
soft-thresholding operator defined as: 
\[
	\cS_{\lambda}(x)=\sgn(x)\cdot\max\{|x|-\lambda, 0\}.  
\]

Subproblem~\eqref{eq:kstep} is a quadratic programming problem
with a simplex constraint. While in principle, it may be solved using iterative
numerical algorithms, in the blind deblurring literature an approximation
scheme is often adopted. The unconstrained quadratic programming problem is
solved first (again using DFT) to obtain a solution; its negative coefficients
are then thresholded out, and finally normalized to have unit sum (Steps 8--10 of Algorithm~\ref{alg:half_quadratic}). We define
${[x]}_+=\max\{x,0\}$. This function is commonly called the Rectified Linear Unit
(ReLU) in neural network terminology~\cite{nair_rectified_2010}. 
Note that in Step~\ref{step:kthresh} of Algorithm~\ref{alg:half_quadratic}, we are adopting a common practice~\cite{cho_fast_2009,pan_deblurring_2018} by thresholding the kernel coefficients using a positive constant (which is usually set as a constant parameter multiplying the maximum of the kernel coefficients); to avoid the non-smoothness of the maximum operation, we use the log-sum-exp function as a smooth surrogate.

We note that the quality of the outputs
and the convergence speed depend crucially on the filters
${\{\mathbf{f}^l_i\}}_{i,l}$ and parameters
${\{\zeta_i^l,\lambda_i^l\}}_{i,l}$, which are difficult to infer due to the
huge variety of real world data.  Under traditional settings, they are usually
determined by hand crafting or domain knowledge.  For example,
in~\cite{perrone_clearer_2016,pan_deblurring_2018} ${\{\mathbf{f}_i^l\}}_{i,l}$
are taken as Prewitt filters while ${\{\lambda_i^l\}}_l$'s are chosen as
geometric sequences. Their optimal values thus remain unclear. To optimize the
performance, we learn (adapt) the filters and parameters by using training
images in a deep learning set-up via back-propagation. A detailed visual
comparison between filters commonly used in conventional algorithms and filters
learned through real datasets by DUBLID is provided in
Section~\ref{subsec:ablation}. 

\begin{algorithm}[H]
	\renewcommand{\algorithmicrequire}{\textbf{Input:}}
	\renewcommand{\algorithmicensure}{\textbf{Output:}}
	\caption{Half-quadratic Splitting Algorithm for Blind Deblurring with Continuation}

	\begin{algorithmic}[1]
		\REQUIRE{Blurred image $\by$, filter banks ${\{\mathbf{f}^l_i\}}_{i,l}$}, positive constant parameters ${\{\zeta^l_i,\lambda^l_i\}}_{i,l}$, number of iterations\footnotemark $L$ and $\varepsilon$.\label{step:init}
		\ENSURE{Estimated kernel $\widetilde{\bk}$, estimated feature maps ${\{\widetilde{\bg_i}\}}_{i=1}^C$.}
		\STATE{Initialize $\bk^1\gets\delta$;\,$\bz^1_i\gets 0,i=1,\dots,C$.}
		\FOR{$l=1$ \TO $L$}
			\FOR{$i=1$ \TO $C$}
			\STATE{$\by_i^l\gets\mathbf{f}^l_i\ast\by$,}\label{step:yfilter}
				\STATE{$\bg_i^{l+1}\gets\cF^{-1}\left\{\frac{\zeta^l_i\widehat{\bk^l}^\ast\odot\widehat{\by^l_i}+\widehat{\bz_i^l}}{\zeta^l_i\left|\widehat{\bk^l}\right|^2+1}\right\}$,}\label{step:gupdate}
				\STATE{$\bz_i^{l+1}\gets\cS_{\lambda^l_i\zeta^l_i}\left\{\bg_i^{l+1}\right\}$,}\label{step:threshold}
			\ENDFOR
			\STATE{$\bk^{l+\frac{1}{3}}\gets\cF^{-1}\left\{\frac{\sum_{i=1}^C\widehat{\bz_i^{l+1}}^\ast\odot\widehat{\by^l_i}}{\sum_{i=1}^C\left|\widehat{\bz_i^{l+1}}\right|^2+\epsilon}\right\}$,}\label{step:kupdate}
			\STATE{$\bk^{l+\frac{2}{3}}\gets{\left[\bk^{l+\frac{1}{3}} - \beta^l\log\left(\sum_i\exp\left(\bk^{l+\frac{1}{3}}_i\right)\right)\right]}_+$,}\label{step:kthresh}
			\STATE{$\bk^{l+1}\gets\frac{\bk^{l+\frac{2}{3}}}{\left\|\bk^{l+\frac{2}{3}}\right\|_1}$,}
			\STATE{$l\gets l+1$.}
		\ENDFOR
	\end{algorithmic}\label{alg:half_quadratic}
\end{algorithm}

\footnotetext{$L$ refers to the number of outer iterations, which subsequently becomes the number of network layers in Section~\ref{subsec:unrolling}. While in traditional iterative algorithms it is commonly determined by certain termination criteria, this approach is difficult to implement for neural networks. Therefore, in this work we choose it through cross-validation as is done in ~\cite{gregor_learning_2010,wang_deep_2015}.}

After Algorithm~\ref{alg:half_quadratic} converges, we obtain the
estimated feature maps ${\{\widetilde{\bg_i}\}}_i$ and the estimated
kernel $\widetilde{\bk}$. Because of the low-pass nature of
$\widetilde{\bk}$, using it alone is inadequate to reliably recover
the sharp image $\bx$ and regularization is needed. We may infer
from~\eqref{eq:observation_model} that, as $\widetilde{\bk}$
approximates $\bk$, $\widetilde{\bg_i}$ should approximate
$\mathbf{f}_i\ast\bx$. Therefore, we retrieve $\bx$ by solving the following
optimization problem:
\begin{align}
	\widetilde{\bx}&\gets\argmin_{\bx}\frac{1}{2}\left\|\by-\widetilde{\bk}\ast\bx\right\|_2^2+\sum_{i=1}^C\frac{\eta_i}{2}\left\|{\mathbf{f}}^L_i\ast\bx-\widetilde{\bg_i}\right\|_2^2\nonumber\\
				   &=\cF^{-1}\left\{\frac{\widehat{\widetilde{\bk}}^\ast\odot\widehat{\by}+\sum_{i=1}^C\eta_i\widehat{{\mathbf{f}}^L_i}^\ast\odot\widehat{\widetilde{\bg_i}}}{\left|\widehat{\widetilde{\bk}}\right|^2+\sum_{i=1}^C\eta_i\left|\widehat{{\mathbf{f}}^L_i}\right|^2}\right\},\label{eq:image_reconstruction}
\end{align}
where $\eta_i$'s are positive regularization parameters.

\section{Algorithm Unrolling for Deep Blind Image Deblurring (DUBLID)}\label{sec:implementation}

\begin{figure}
	\setcounter{subfigure}{0}
	\subfloat[\label{subfig:feature}] {\includegraphics[height=0.21\textheight]{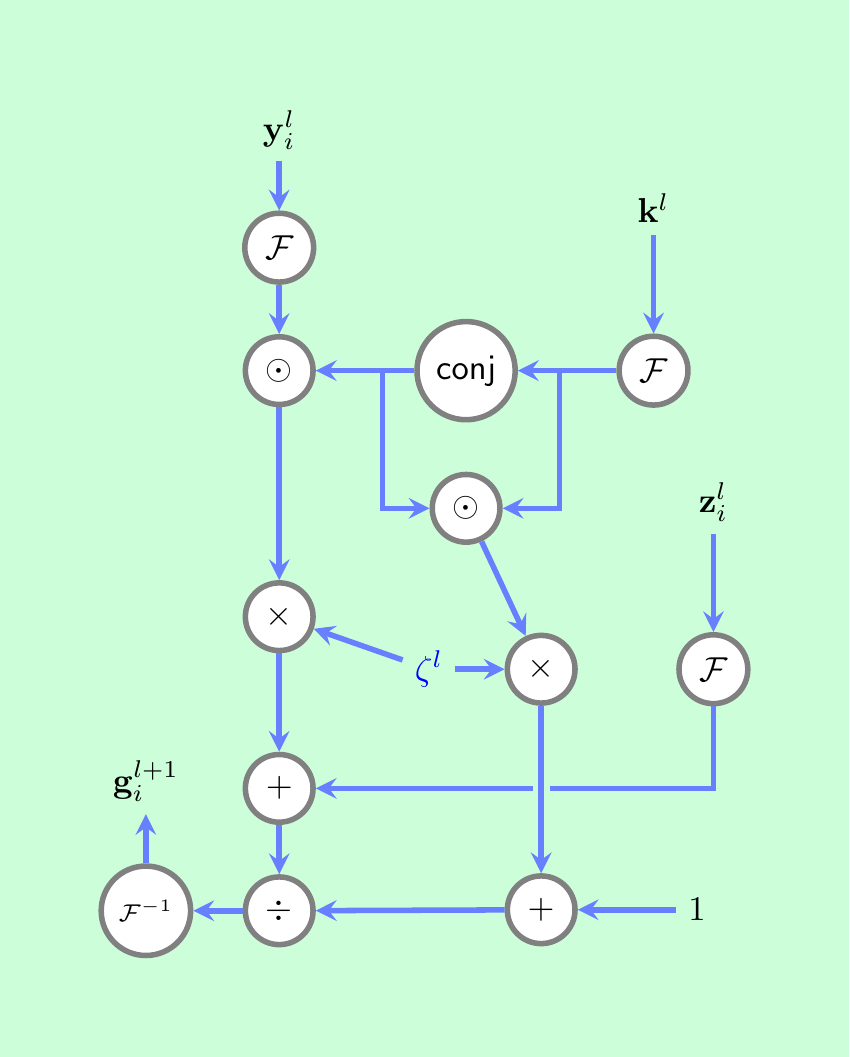}}\,\!
	\subfloat[\label{subfig:kernel}] {\includegraphics[height=0.21\textheight]{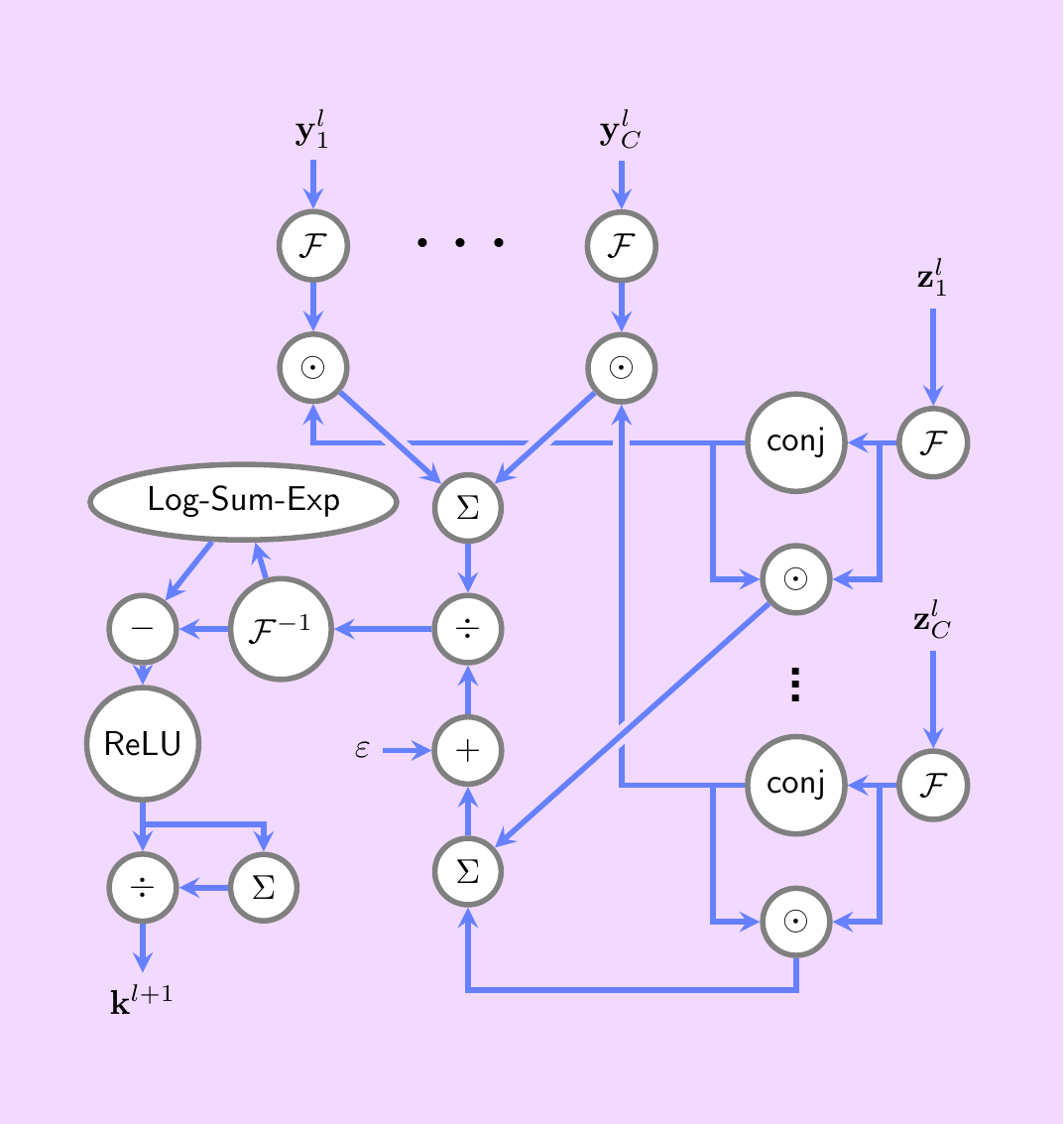}}\\
	\subfloat[\label{subfig:image}] {\includegraphics[width=\linewidth]{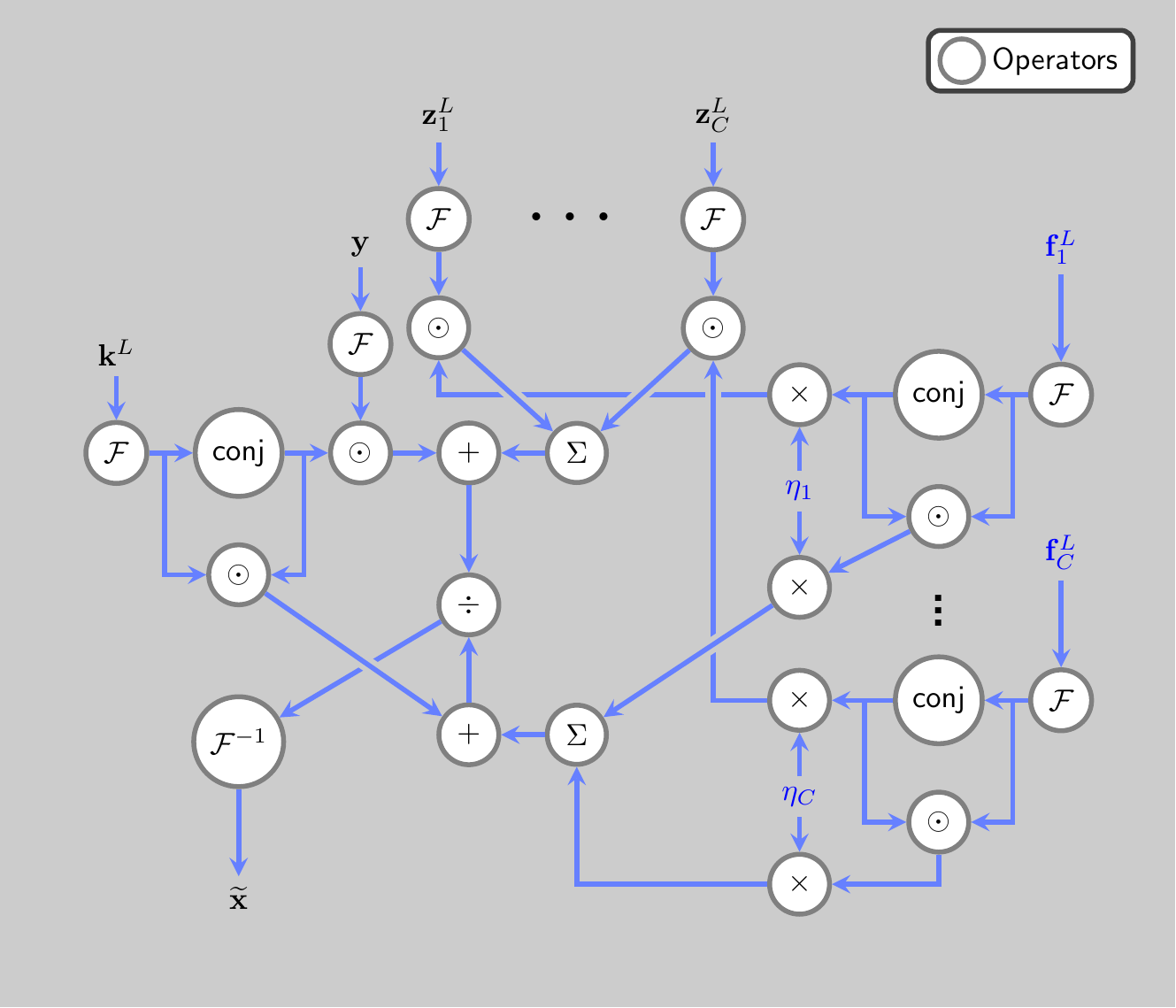}}
	\caption{Block diagram representations of \protect\subref{subfig:feature} step~\ref{step:gupdate} in Algorithm~\ref{alg:half_quadratic}, \protect\subref{subfig:kernel} step~\ref{step:kupdate} in Algorithm~\ref{alg:half_quadratic} and \protect\subref{subfig:image} Equation~\eqref{eq:image_reconstruction}. After unrolling the iterative algorithm to form a multi-layer network, the diagramatic representations serve as building blocks that repeat themselves from layer to layer. The parameters ($\zeta$ and $\eta$) are learned from real datasets and colored in blue.}\label{fig:diagrams}\vspace{-10pt}
\end{figure}

\subsection{Network Construction via Algorithm Unrolling}\label{subsec:unrolling}
\begin{figure*}
	\centering\includegraphics[width=0.95\textwidth]{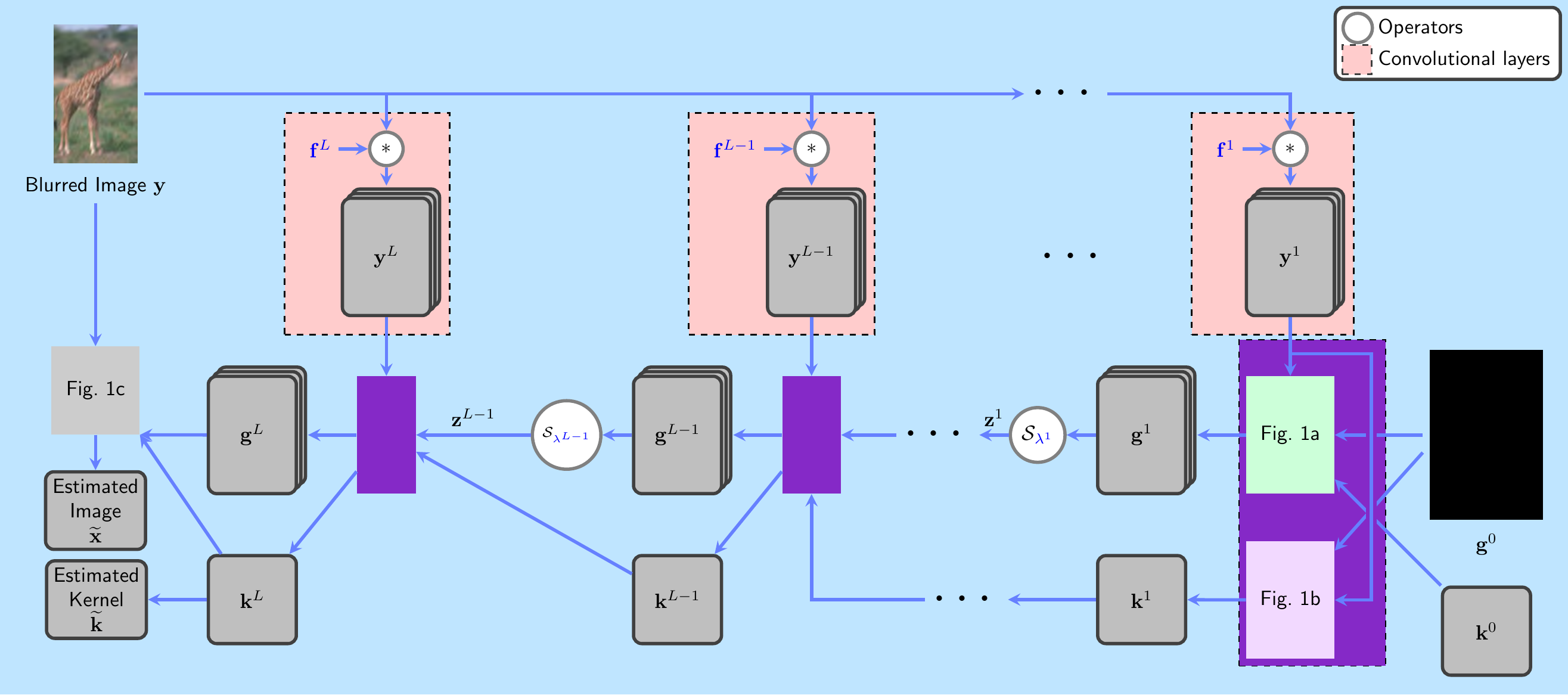}\vspace{-5pt}
	\caption{Algorithm~\ref{alg:half_quadratic} unrolled as a neural network. The parameters that are learned from real datasets are colored in blue.}\label{fig:unroll}\vspace{-8pt}
\end{figure*}

Each step of Algorithm~\ref{alg:half_quadratic} is in analytic form
and can be implemented using a series of basic functional operations.
In particular, step~\ref{step:gupdate} and step~\ref{step:kupdate} in
Algorithm~\ref{alg:half_quadratic} can be implemented according to
the diagrams in Fig.~\ref{subfig:feature} and
Fig.~\ref{subfig:kernel}, respectively. The soft-thresholding operation in step~\ref{step:threshold} may be implemented using two ReLU operations by recognizing that $\cS_{\lambda}(x)={[x-\lambda]}_+-{[-x-\lambda]}_+$.
Similarly,~\eqref{eq:image_reconstruction} may be implemented
according to Fig.~\ref{subfig:image}. Therefore, each iteration of
Algorithm~\ref{alg:half_quadratic} admits a diagram representation,
and repeating it $L$ times yields an $L$-layer neural network (as
shown in Fig.~\ref{fig:unroll}) which corresponds to executing
Algorithm~\ref{alg:half_quadratic} with $L$ iterations.  For
notational brevity, we concatenate the parameters in each layer and
let
$\mathbf{f}^l={(\mathbf{f}_i^l)}_{i=1}^C,\zeta^l={(\zeta_i^l)}_{i=1}^C,\lambda^l={(\lambda_i^l)}_{i=1}^C$
and $\eta={(\eta_i)}_{i=1}^C$. We also concatenate $\by_i^l$'s,
$\bz_i^l$'s and $\bg_i^l$'s by letting $\by^l={(\by_i^l)}_{i=1}^C$,
$\bz^l={(\bz_i^l)}_{i=1}^C$ and $\bg^l={(\bg_i^l)}_{i=1}^C$,
respectively.

When the blur kernel has a large size (which may happen due to fast motion), it is desirable to alter the spatial size of the
filter banks ${\{\mathbf{f}_i\}}_i$ in different layers. In blind deblurring,
kernel recovery is frequently performed stage-wise in a coarse-to-fine scheme:
the kernel is initially estimated from a high-level summary of the image
features, and then progressively refined by introducing lower-level details.
For example, in~\cite{xu_two-phase_2010} an initial kernel is obtained by
masking out detrimental gradients, followed by iterative refinements. Other
works~\cite{cho_fast_2009,perrone_clearer_2016,pan_deblurring_2018} use a
multi-scale pyramid implementation by decomposing the images into a Gaussian
pyramid and progressively refine the kernel in each scale. We may integrate
this scheme into Algorithm~\ref{alg:half_quadratic} by choosing large filters
in early iterations, so that they are capable of capturing rather high-level
features, and gradually decrease their size to let fine details emerge.
Translating this into the network, we may expect the following
relationship among the sizes of kernels in different layers:
\[
	\text{size of }\mathbf{f}_i^1\geq\text{ size of }\mathbf{f}_i^2\geq\text{ size of }\mathbf{f}_i^3\geq\dots.
\]
In practice, large filters may be difficult to train due to the large number of parameters they contain. To address this issue, we produce large filters by cascading small $3\times 3$ filters, following the same principle as~\cite{simonyan_very_2015}. Formally, we set $\mathbf{f}_i^L=\bw_{i1}^L$ where ${\{\bw_{i1}^L\}}_{i=1}^C$ is a collection of $3\times 3$ filters, and recursively obtain $\mathbf{f}_i^l$ by filtering $\mathbf{f}_i^{l+1}$ through $3\times 3$ filters ${\{\bw_{ij}^l\}}_{i,j=1}^C$:
\[
	\mathbf{f}_i^l\gets\sum_{j=1}^C\bw_{ij}^l\ast\mathbf{f}_j^{l+1},\quad i=1,2,\dots,C.
\]

\noindent Embedding the above into the network, we obtain the structure
depicted in Fig.~\ref{fig:network}. Note that $\by^l$ can now be
obtained more efficiently by filtering $\by^{l+1}$ through $\bw^l$. Also note that ${\{\bw_{ij}^l\}}_{i,j=1}^C$ are to be learned as marked in Fig.~\ref{fig:network}. Experimental justification of cascaded filtering is provided in Fig.~\ref{fig:cascade}.

\subsection{Training}\label{subsec:train}
In a given training set, for each blurred image
$\by_t^\mathrm{train}(t=1,\dots,T)$, we let the corresponding sharp image and
kernel be $\bx_t^\mathrm{train}$ and $\bk_t^\mathrm{train}$, respectively. We
do not train the parameter $\varepsilon$ in step~\ref{step:kupdate} of
Algorithm~\ref{alg:half_quadratic} because it simply serves as a small constant
to avoid division by zeros. We re-parametrize $\zeta_i^l$ in
step~\ref{step:threshold} of Algorithm~\ref{alg:half_quadratic} by letting
$b_i^l=\lambda_i^l\zeta_i^l$ and denote $b^l={(b_i^l)}_{i=1}^C,l=1,\dots,L$.  The
network outputs $\widetilde{\bx}_t,\widetilde{\bk}_t$ corresponding to
$\by^\mathrm{train}_t$ depend on the parameters $\bw^l$, $b^l,\zeta^l,\beta^l$,
$l=1,2,\dots,L$.  In addition, $\widetilde{\bx}_t$ depends on $\eta$. We train the
network to determine these parameters by solving the following optimization problem:
\begin{align}
	\min_{{\{\bw^l,b^l,\zeta^l,\beta^l\}}_{l},\eta,{\{\tau_t\}}_t}&\resizebox{.7\linewidth}{!}{$\sum_{t=1}^T\frac{\kappa_t}{2}\mse\left(\widetilde{\bk}_t\left({\left\{\bw^l,b^l,\zeta^l,\beta^l\right\}}_{l}\right),\cT_{\tau_t}\left\{\bk_t^{\mathrm{train}}\right\}\right)$}\nonumber\\
	+\frac{1}{2}&\resizebox{.7\linewidth}{!}{$\mse\left(\widetilde{\bx}_t\left({\{\bw^l,b^l,\zeta^l,\beta^l\}}_{l},\eta\right),\cT_{-\tau_t}\left\{\bx_t^{\mathrm{train}}\right\}\right)\label{eqn:loss}$}\nonumber\\
	\resizebox{.32\linewidth}{!}{$\text{subject to }b_i^l\geq0, \lambda_i^l\geq 0,$}&\resizebox{.52\linewidth}{!}{$\beta^l\geq 0,\quad l=1,\dots,L,i=1,\dots,C,$}
\end{align}
where $\kappa_t>0$ is a constant parameter which is fixed to $\frac{\kappa_0}{\max_i|{(\bk_t^{\mathrm{train}})}_i|^2}$, and we determined $\kappa_0=10^5$ through cross-validation. $\mse(\cdot,\cdot)$ is the (empirical) Mean-Square-Error loss function, and $\cT_{\tau}\left\{\cdot\right\}$ is the translation operator in 2D that performs a shift by $\tau\in\mathbb{R}^2$. The shift operation is used here to compensate for the inherent shifting ambiguity in blind deconvolution~\cite{perrone_clearer_2016}.

\begin{figure*}
	\centering
	\includegraphics[width=.95\textwidth]{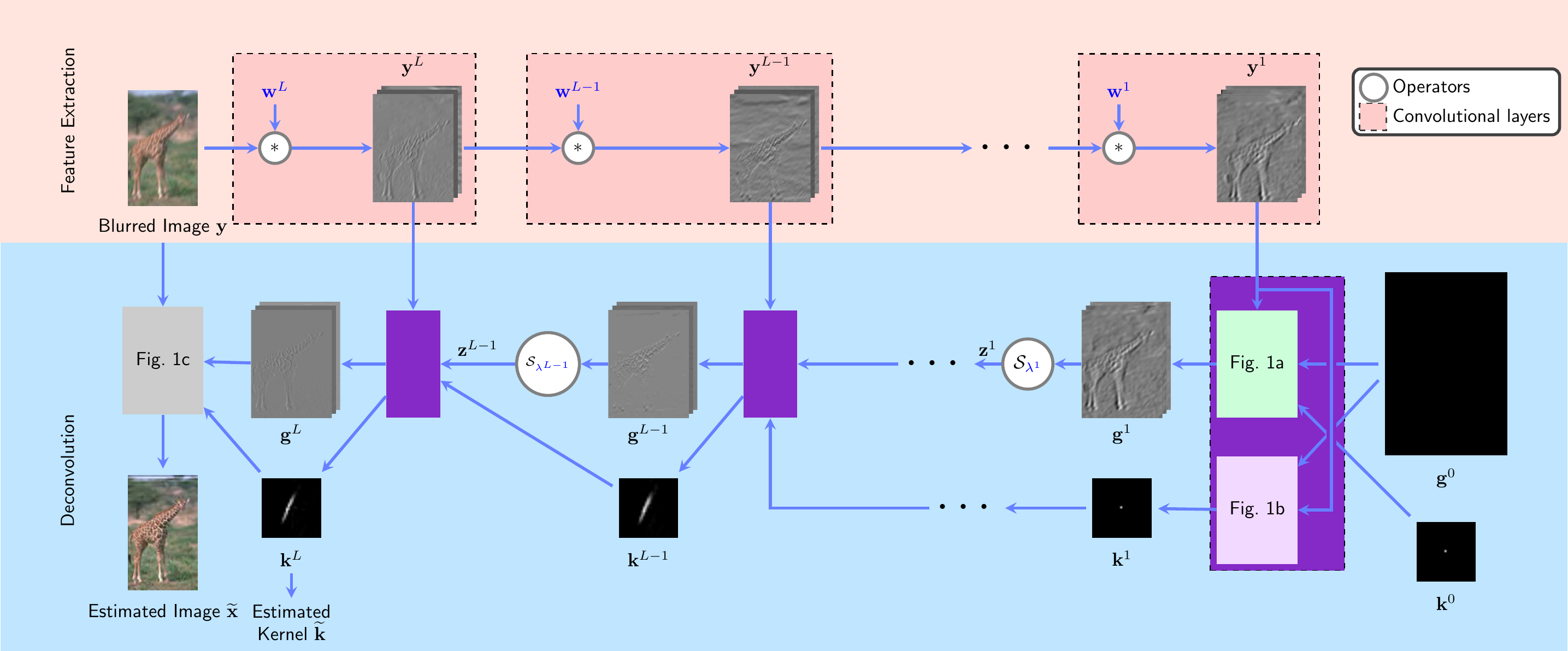}\vspace{-5pt}
	\caption{DUBLID: using a cascade of small $3\times 3$ filters instead of large filters (as compared to the network in Fig.~\ref{fig:unroll}) reduces the dimensionality of the parameter space, and the network can be easier to train. Intermediate data (hidden layers) on the trained network are also shown. It can be observed that, as $l$ increases, $\bg^l$ and $\by^l$ evolve in a coarse-to-fine manner. The parameters that will be learned from real datasets are colored in blue.\vspace{-10pt}}\label{fig:network}
\end{figure*}

In the training process, when working on each mini-batch, we alternate between
minimizing over ${\{\tau_t\}}_t$ and performing a projected stochastic gradient
descent step. Specifically, we first determine the optimal ${\{\tau_t\}}_t$
efficiently via a grid search in the Fourier domain.  We then take one
stochastic gradient descent step; the analytic derivation of the gradient is
included in Appendix~\ref{sec:backpropagation}. Finally, we threshold out the
negative coefficients of ${\{b_i^l,\zeta_i^l,\beta^l\}}_{i,l}$ to enforce the
non-negativity constraints.  We use the Adam algorithm~\cite{kingma_adam:_2015}
to accelerate the training speed.  The learning rate is set to $1\times
10^{-3}$ initially and decayed by a factor of $0.5$ every $20$ epochs.  We
terminate training after $160$ epochs. The parameters ${\{b_i^l\}}_{i,l}$ are
initialized to $0.02$, ${\{\zeta_i^l\}}_{i,l}$ initialized to $1$, ${\{\beta^l\}}_l$ initialized to $0$, and
${\{\eta_i\}}_i$ initialized to $20$, respectively. These values are again
determined through cross-validation. The upper part (feature extraction
portion) of the network in Fig.~\ref{fig:network} resembles a CNN with linear
activations (identities) and thus we initialize the weights according
to~\cite{glorot_understanding_2010}.\vspace{-10pt}

\subsection{Handling Color Images}\label{subsec:color}
For color images, the red, green and blue channels $\by_r$, $\by_g$, and $\by_b$ are blurred by the same kernel, and thus the following model holds instead of~\eqref{eq:blur_model}: 
\[
\by_c=\bk\ast\bx_c+\bn_c,\quad c\in\{r,g,b\}.
\]

To be consistent with existing literature, we modify $\bw^L$ in
Fig.~\ref{fig:network} to allow for multi-channel inputs. More
specifically, $\by^L$ is produced by the following formula:
\[
	\by^L_i=\sum_{c\in\{r,g,b\}}\bw^L_{ic}\ast\by_c,\quad i=1,\dots,C.
\]\vspace{-3pt}

It is easy to check that, with $\bw^L$ and $\by^L$ being replaced,
all the components of the network can be left unchanged except for
the module in Fig.~\ref{subfig:image}. This is
because~\eqref{eq:observation_model} no longer holds
and is modified to the following:\vspace{-3pt}
\[
	\sum_{c\in\{r,g,b\}}\bw_{ic}\ast\by_c=\bk\ast\left(\sum_{c\in\{r,g,b\}}\bw_{ic}\ast\bx_c\right)+\bn_i^\prime,\,i=1,\dots,C,
\]\vspace{-0pt}
where $\bn_i^\prime=\sum_{c\in\{r,g,b\}}\bw_{ic}\ast\bn$ represents
Gaussian noise.

Problem~\eqref{eq:image_reconstruction} then becomes:\vspace{-3pt}
\begin{align}
	\{\widetilde{\bx_r},\widetilde{\bx_g},\widetilde{\bx_b}\}&\gets\argmin_{\bx_r,\bx_g,\bx_b}\sum_{c\in\{r,g,b\}}\frac{1}{2}\left\|\by_c-\widetilde{\bk}\ast\bx_c\right\|_2^2\nonumber\\
															 &+\sum_{i=1}^C\frac{\eta_i}{2}\left\|\sum_{c\in\{r,g,b\}}\bw_{ic}\ast\bx_c-\widetilde{\bg_i}\right\|_2^2,\nonumber
\end{align}\vspace{-3pt}
whose solution is given as follows:\vspace{-3pt}
\begin{align*}
	\widetilde{\bx_r}&=\cF^{-1}\left\{\frac{\bm_{rr}\odot\bb_r+\bm_{rg}\odot\bb_g+\bm_{rb}\odot\bb_b}{\bd}\right\},\\
	\widetilde{\bx_g}&=\cF^{-1}\left\{\frac{\bm_{rg}^\ast\odot\bb_r+\bm_{gg}\odot\bb_g+\bm_{gb}\odot\bb_b}{\bd}\right\},\\
	\widetilde{\bx_b}&=\cF^{-1}\left\{\frac{\bm_{rb}^\ast\odot\bb_r+\bm_{gb}^\ast\odot\bb_g+\bm_{bb}\odot\bb_b}{\bd}\right\},
\end{align*}\vspace{-3pt}
where\vspace{-3pt}
\begin{align*}
	\bm_{rr}&=\bc_{gg}\odot\bc_{bb}-\left|\bc_{gb}\right|^2,\,\bm_{rg}=\bc_{rb}\odot\bc_{gb}^\ast-\bc_{bb}\odot\bc_{rg},\\
	\bm_{rb}&=\bc_{rg}\odot\bc_{gb}-\bc_{gg}\odot\bc_{rb},\,\bm_{gg}=\bc_{bb}\odot\bc_{rr}-\left|\bc_{rb}\right|^2,\\
	\bm_{gb}&=\bc_{rg}^\ast\odot\bc_{rb}-\bc_{rr}\odot\bc_{gb},\,\bm_{bb}=\bc_{rr}\odot\bc_{gg}-\left|\bc_{rg}\right|^2.
\end{align*}\vspace{-3pt}
Here,\vspace{-3pt}
\begin{align*}
	\bc_{cc^\prime}&=\sum_{i=1}^C\eta_i\widehat{\bw_{ic}}^\ast\odot\widehat{\bw_{ic^\prime}}+\left|\widehat{\bk}\right|^2\delta_{cc^\prime},\quad c,c'\in\{r,g,b\},\\\vspace{-1pt}
	\bb_c&=\widehat{\bk}^\ast\odot\widehat{\by_c}+\sum_{i=1}^C\eta_i\widehat{\bw_{ic}}^\ast\odot\widehat{\bg_i},\quad c\in\{r,g,b\},\\\vspace{-1pt}
	\bd&=\left(\bc_{gg}\odot\bc_{rr}-\left|\bc_{rg}\right|^2\right)\odot\bc_{bb}+2\Re\{\bc_{rb}^\ast\odot\bc_{rg}\odot\bc_{gb}\}\\
	   &-\left|\bc_{gb}\right|^2\odot\bc_{rr}-\left|\bc_{rb}\right|^2\odot\bc_{gg},
\end{align*}\vspace{0pt}
and $\delta_{cc^\prime}$ is the Kronecker delta function.
These analytical formulas may be represented using
diagrams similar to Fig.~\ref{subfig:image} and embedded into a
network.\vspace{-5pt}

\section{Experimental Verification}\label{sec:experiments}
\subsection{Experimental Setups}\label{subsec:setup}
\subsubsection{Datasets, Training and Test Setup}
\begin{itemize}
	\item \textbf{Training for linear kernels}: For the images we used the Berkeley Segmentation Data Set 500 (BSDS500)~\cite{arbelaez_contour_2011} which is a large dataset of $500$ natural images that is explicitly divided into disjoint training, validation and test subsets. Here we use $300$ images for training by combining the training and validation images. \\
	We generated 256 linear kernels by varying the length and angle of the kernels (refer to Section~\ref{subsec:linear} for details).
	\item \textbf{Training for nonlinear kernels}: We used the Microsoft COCO~\cite{lin2014microsoft} dataset which is a large-scale object detection, segmentation, and captioning dataset containing 330K images. \\
	\textit{Nonlinear kernels}: We generated around 30,000 real world kernels by recording camera motion trajectories (refer to Sec. \ref{subsec:nonlinear} for details).
	\item \textbf{Testing for the linear kernel experiments}: We use $200$ images from the test portion of BSDS500 as test images and we randomly choose four kernels of different angle and length as test kernels.
	\item \textbf{Testing for the nonlinear kernel experiments:} We test on {\em two} benchmark datasets specifically developed for evaluating blind deblurring with non-linear kernels: 1.)	4 images and 8 kernels by Levin \textit{et al.}~\cite{levin_understanding_2011} and 2.) 80 images and 8 kernels from Sun {\it et al.}~\cite{sun_edge-based_2013}.
\end{itemize}

\subsubsection{Comparisons Against State of the Art Methods} We compare against \textbf{five} methods:
\begin{itemize}
	\item Perrone {\it et al.}~\cite{perrone_clearer_2016} - a representative iterative blind image deblurring method based on total-variation minimization, which demonstrated state-of-the-art performance amongst traditional iterative methods. (TPAMI 2016)
	\item Chakrabarti {\it et al.}~\cite{chakrabarti_neural_2016} - one of the first neural network blind image deblurring methods. (CVPR 2016)
	\item Nah {\it et al.}~\cite{nah_deep_2017} - a recent deep learning method based on the state-of-the-art ResNet~\cite{he_deep_2016}. (CVPR 2017)
	\item Kupyn {\it et al.}~\cite{kupyn_deblurgan_2018} - a recent deep learning method based on the state-of-the-art generative adversarial networks (GAN)~\cite{goodfellow_generative_2014}. (CVPR 2018)
	\item Xu {\it et al.}~\cite{xu_motion_2018} - a recent state of the art deep learning method focused on motion kernel estimation. (TIP 2018)
\end{itemize}

\subsubsection{Quantitative Performance Measures} To quantitatively assess the performance of various methods in different scenarios, we use the following metrics:
\begin{itemize}
	\item Peak-Signal-to-Noise Ratio (PSNR);
	\item Improvement in Signal-to-Noise-Ratio (ISNR), which is given by $ \mathrm{ISNR}=10\log_{10}\left(\frac{\|\by-\bx\|_2^2}{\|\widetilde{\bx}-\bx\|_2^2}\right)$
		where $\widetilde{\bx}$ is the reconstructed image;
	\item Structural Similarity Index (SSIM)~\cite{wang_image_2004};
	\item (Empirical) Root-Mean-Square Error (RMSE) computed between the estimated kernel and the ground truth kernel.
\end{itemize}

We note that for selected methods, RMSE numbers (against the ground truth kernel) are {\em not\/} reported in Tables~\ref{tab:linear},~\ref{tab:levin} and~\ref{tab:sun} because those methods directly estimate just the deblurred image (and not the blur kernel).\vspace{-5pt}

\subsection{Ablation Study of the Network}\label{subsec:ablation}

To provide insights into the design of our network, we first carry out a series
of experimental studies to investigate the influence of two key design variables: 1.) the number of layers $L$, and 2.) the number of filters $C$. We monitor the performance
using PSNR and RMSE. The results in Tables~\ref{tab:layer_number} and ~\ref{tab:channel_number} are for linear kernels with the same training-test configuration as in Section \ref{subsec:setup}. The trends over $L$ and $C$ were found to be similar for non-linear kernels as well.

We first study the influence of different number of layers
$L$. We alter $L$, i.e. the proposed DUBLID is learned as in Section~\ref{subsec:train}, each
time as $L$ is varied. Table~\ref{tab:layer_number} summarizes the numerical
scores corresponding to different $L$. Clearly, the network performs better
with increasing $L$, consistent with common
observations~\cite{kim_accurate_2016,he_deep_2016}. In addition, the network
performance improves marginally after $L$ reaches $10$. We thus fix $L=10$
subsequently. For all results in Table~\ref{tab:layer_number}, the number of
filters $C$ is fixed to $16$.

\begin{table}
	\centering
	\caption{Effects of different values of layer $L$.}\label{tab:layer_number}
	\begin{tabular}{ccccc}
		\toprule
		Number of layers & $6$ & $8$ & $10$ & $12$\\
		\midrule
		PSNR (dB) & $26.55$ & $26.94$ & $27.30$ & $27.35$\\
		\midrule
		RMSE ($\times 10^{-3}$) & $1.96$ & $2.06$ & $1.67$ & $1.66$\\
		\bottomrule
	\end{tabular}\vspace{-8pt}
\end{table}

We next study the effects of different values of $C$ in a similar fashion. The network performance over different choices of $C$ is summarized in Table~\ref{tab:channel_number}. It can be seen that the network performance clearly improves as $C$ increases from $8$ to $16$. However, the performance slightly drops when $C$ increases further, presumably because of overfitting. We thus fix $C=16$ henceforth.

\begin{table}
	\centering
	\caption{Effects of different values of number of filters $C$.}\label{tab:channel_number}
	\begin{tabular}{cccc}
		\toprule
		Number of filters & $8$ & $16$ & $32$ \\
		\midrule
		PSNR (dB) & $26.55$ & $27.30$ & $27.16$ \\
		\midrule
		RMSE ($\times 10^{-3}$) & $1.99$ & $1.67$ & $1.93$ \\
		\bottomrule
	\end{tabular}\vspace{-8pt}
\end{table}

\begin{figure}
	\centering{\subfloat[\label{subfig:learned} DUBLID-Learned]{\includegraphics[width=0.95\linewidth]{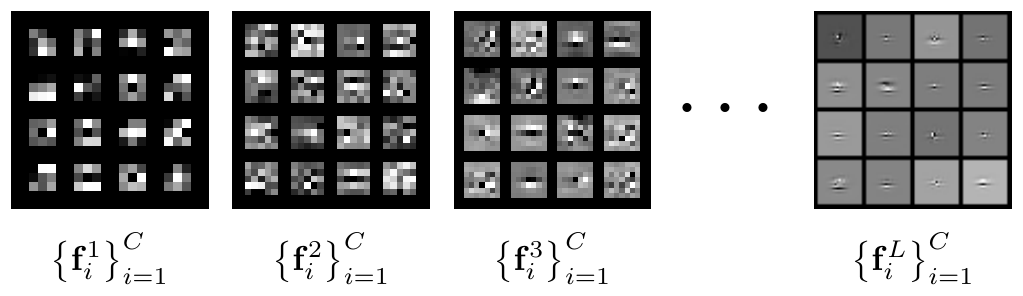}}\,\!}\\\vspace{-5pt}
	\subfloat[\label{subfig:analytic}DUBLID-Sobel]{\includegraphics[width=0.25\linewidth]{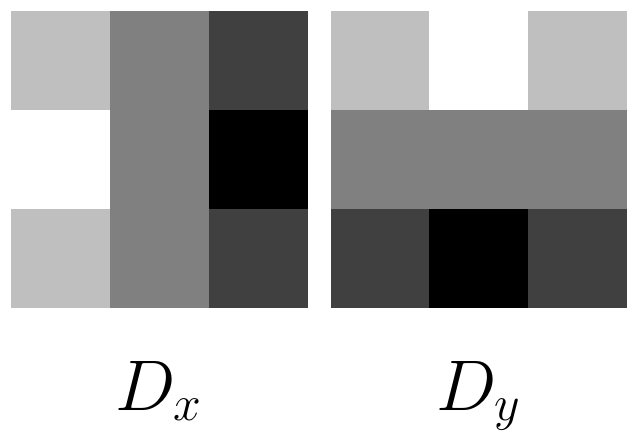}}\\\vspace{-5pt}
	\subfloat[\label{subfig:true_kernel}]{\includegraphics[width=0.28\linewidth]{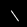}}\,\!
	\subfloat[\label{subfig:kernel_learned}]{\includegraphics[width=0.28\linewidth]{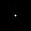}}\,\!
	\subfloat[\label{subfig:kernel_tv}]{\includegraphics[width=0.28\linewidth]{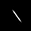}}\\
	\caption{Comparison of learned filters with analytic Sobel filters: \protect\subref{subfig:learned} DUBLID learned filters. \protect\subref{subfig:analytic} Sobel filters that are commonly employed in traditional iterative blind deblurring algorithms. \protect\subref{subfig:true_kernel} An example motion blur kernel. \protect\subref{subfig:kernel_learned} Reconstructed kernel using Sobel filters and \protect\subref{subfig:kernel_tv} using learned filters.}\label{fig:learned_filters}\vspace{-10pt}
\end{figure}

To corroborate the network design choices made in
Section~\ref{subsec:unrolling}, we illustrate DUBLID performance for different filter choices. We first verify the significance of learning the filters
${\{\bw^l\}}_l$ (and in turn ${\{\mathbf{f}_i^l\}}_{i,l}$) and compare the
performance with a typical choice of analytical filters, the Sobel filters, in
Fig.~\ref{fig:learned_filters}. Note that by employing Sobel filters, the
network reduces to executing TV-based deblurring but for a small number of iterations, which coincides with the number of layers $L$. For
fairness in comparison, the fixed Sobel filter version of DUBLID (called DUBLID-Sobel) is trained exactly as in Section III-B to optimize other parameters. As Fig.~\ref{fig:learned_filters} reveals, DUBLID-Sobel is unable to accurately recover the kernel. Indeed, such phenomenon has been
observed and analytically studied by Levin {\it et
al.}~\cite{levin_understanding_2011}, where they point out that traditional
gradient-based approaches can easily get stuck at a delta solution. To gain
further insight, we visualize the learned filters as well as the Sobel filters
in Fig.~\ref{subfig:learned} and Fig.~\ref{subfig:analytic}. The learned
filters demonstrate richer variations than known analytic (Sobel) filters and
are able to capture higher-level image features as $l$ grows. This enables the
DUBLID network to better recover the kernel coefficients subsequently.
Quantitatively, the PSNR achieved by DUBLID-Sobel for $L=10$ and $C=16$ on the
same training-test set up is $18.60$ dB, which implies
that DUBLID achieves a $8.7$ dB gain by explicitly optimizing filters in a
data-adaptive fashion.

\begin{figure}\centering
	\subfloat[\label{subfig:original_kernel}]{\includegraphics[width=0.28\linewidth]{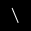}}\,\!
	\subfloat[\label{subfig:fixed}]{\includegraphics[width=0.28\linewidth]{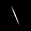}}\,\!
	\subfloat[\label{subfig:cascaded}]{\includegraphics[width=0.28\linewidth]{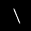}}\\
	\caption{The effectiveness of cascaded filtering: \protect\subref{subfig:original_kernel} a sample motion kernel. \protect\subref{subfig:fixed} Reconstructed kernel by fixing all $\mathbf{f}_i^l$'s to be of size $3\times 3$, which can be implemented by enforcing $\bw_{ij}^l$ to be of size $1\times 1$ whenever $l<L$. \protect\subref{subfig:cascaded} Reconstructed kernel using the cascaded filtering structure in Fig.~\ref{fig:network}.\vspace{-10pt}}\label{fig:cascade}
\end{figure}

Finally, we show the effectiveness of cascaded filtering. To this end, we
compare with the alternative scheme of fixing the size of
${\{\mathbf{f}_i^l\}}_{i,l}$ by restricting ${\{\bw^l\}}_l$ to be of size
$1\times 1$ whenever $l<L$. The results are shown in Fig.~\ref{fig:cascade}. By
employing learnable filters, the network becomes capable of capturing the
correct directions of blur kernels as shown in Fig.~\ref{subfig:fixed}. In the absence of cascaded filtering though, the recovered kernel is still coarse -- a limitation that is overcome by using cascaded filtering, verified in
Fig.~\ref{subfig:cascaded}.\vspace{-5pt}

\subsection{Evaluation on Linear Kernels}\label{subsec:linear}
\begin{figure}
	\centering
	\subfloat[\label{subfig:sharp}] {\includegraphics[width=1.8cm]{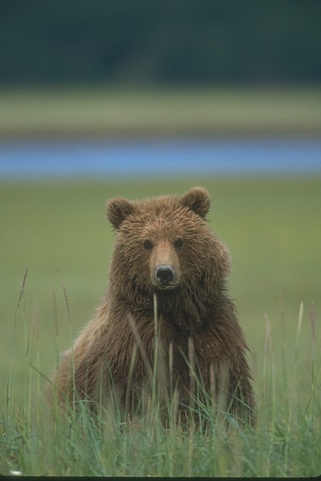}\,\includegraphics[width=1.8cm]{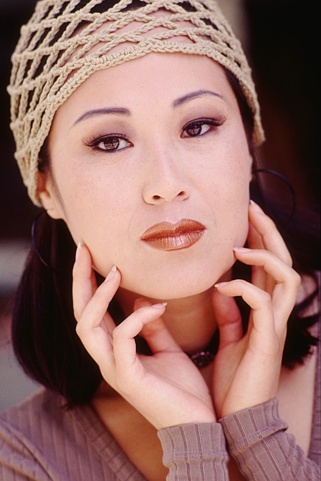}\,\includegraphics[width=1.8cm]{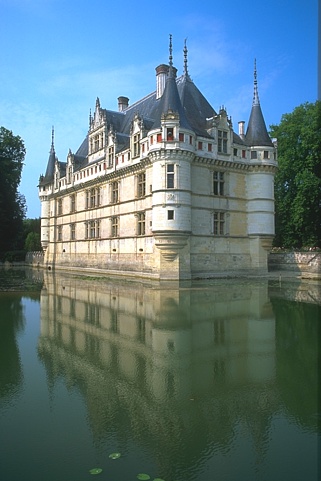}\,\includegraphics[width=1.8cm]{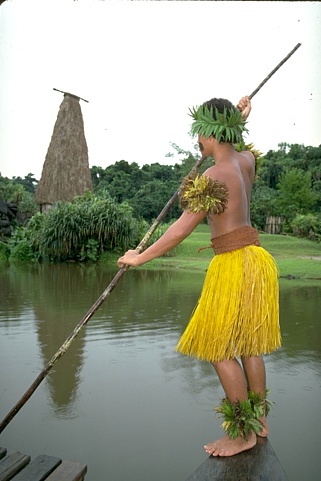}}\\\vspace{-5pt}
	\subfloat[\label{subfig:blur_kernel}] {\includegraphics[width=.9cm]{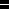}\,\includegraphics[width=.9cm]{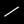}\,\includegraphics[width=.9cm]{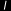}\,\includegraphics[width=.9cm]{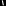}\,\includegraphics[width=.9cm]{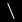}\,\includegraphics[width=.9cm]{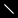}\,\includegraphics[width=.9cm]{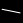}\,\includegraphics[width=.9cm]{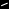}}\\\vspace{-5pt}
	\subfloat[\label{subfig:blurred}] {\includegraphics[width=1.8cm]{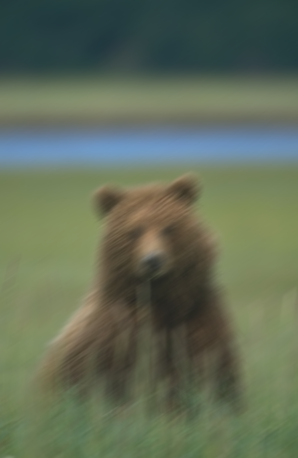}\,\includegraphics[width=1.8cm]{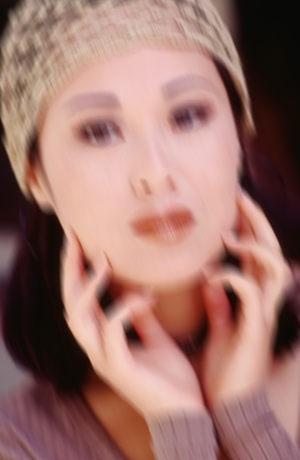}\,\includegraphics[width=1.8cm]{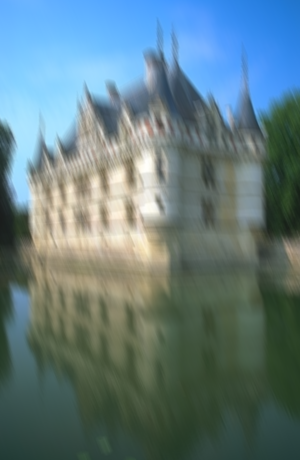}\,\includegraphics[width=1.8cm]{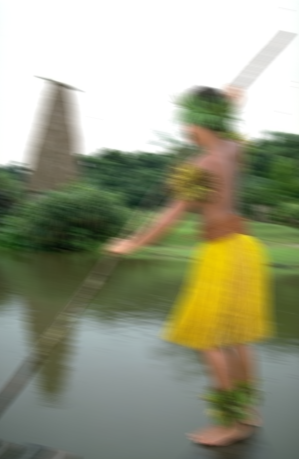}}
	\caption{Examples of images and kernels used for training. \protect\subref{subfig:sharp} The sharp images are collected from the BSDS500 dataset~\cite{martin_database_2001}. \protect\subref{subfig:blur_kernel} The blur kernels are linear motion of different lengths and angles. \protect\subref{subfig:blurred} The blurred images are synthesized by convolving the sharp images with the blur kernels.}\label{fig:train_sample}\vspace{-10pt}
\end{figure}

We use the training and validation portions of the BSDS500~\cite{martin_database_2001,arbelaez_contour_2011} dataset as training images.
The linear motion kernels are generated by uniformly sampling $16$ angles in
$[0,\pi]$ and $16$ lengths in $[5,20]$, followed by 2D spatial interpolation. This gives a total number of $256$
kernels. We then generate $T=256\times 300$ blurred images by convolving each
kernel with each image and adding white Gaussian noise with standard deviation $0.01$ (suppose the image intensity is in range $[0,1]$) individually.
Examples of training samples (images and kernels) are shown in
Fig.~\ref{fig:train_sample}. 
We use $200$ images from the test portion of the BSDS500
dataset~\cite{martin_database_2001} for evaluation. We randomly
choose angles in $[0,\pi]$ and lengths in $[5,20]$ to generate
$4$ test kernels. The images and kernels and convolved to synthesize
$800$ blurred images. White Gaussian noise (again with standard deviation $0.01$) is also added. 

Note that some of the state of the art methods compared against are only
designed to recover the kernels, including~\cite{xu_motion_2018}
and~\cite{perrone_clearer_2016}. To get the deblurred image, the non-blind
method in~\cite{pan_$l_0$_2017} is used consistently. The scores
are averaged and summarized in Table~\ref{tab:linear}. The RMSE values are computed over kernels, and smaller values indicate more accurate recoveries. For all other metrics on images, higher scores generally imply better performance. We do not
include results from Chakrabarti {\it et al.}~\cite{chakrabarti_neural_2016}  here because that method
works on grayscale images only. Table~\ref{tab:linear} confirms that DUBLID
outperforms competing state-of-the art algorithms by a significant margin.  

\begin{figure*}
	\centering
	\subfloat{%
		\begin{tikzpicture}[spy using outlines={rectangle,magnification=8,width=4.2cm,height=2.5cm}]
		\node {\includegraphics[height=0.11\textheight]{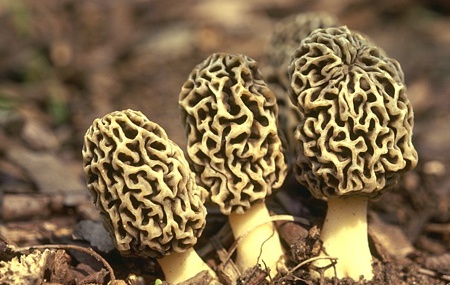}\llap{\includegraphics[height=0.03\textheight]{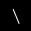}}};
		\spy[blue] on (-0.8,-0.5) in node[below,ultra thick] at (0,-1.5);
		\end{tikzpicture}
	}\hspace{-3mm}
	\subfloat{%
		\begin{tikzpicture}[spy using outlines={rectangle,magnification=8,width=4.2cm,height=2.5cm}]
		\node {\includegraphics[height=0.11\textheight]{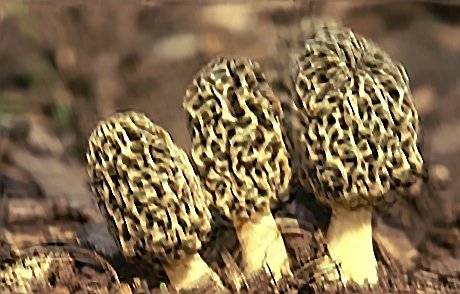}\llap{\includegraphics[height=0.03\textheight]{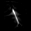}}};
		\spy[blue] on (-0.8,-0.5) in node[below,ultra thick] at (0,-1.5);
		\end{tikzpicture}
	}\hspace{-3mm}
	\subfloat{%
		\begin{tikzpicture}[spy using outlines={rectangle,magnification=8,width=4.2cm,height=2.5cm}]
		\node {\includegraphics[height=0.11\textheight]{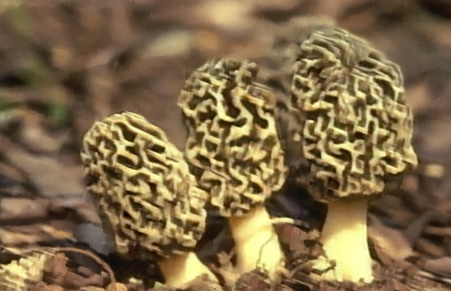}};
		\spy[blue] on (-0.8,-0.5) in node[below,ultra thick] at (0,-1.5);
		\end{tikzpicture}
	}\hspace{-3mm}
	\subfloat{%
		\begin{tikzpicture}[spy using outlines={rectangle,magnification=8,width=4.2cm,height=2.5cm}]
		\node {\includegraphics[height=0.11\textheight]{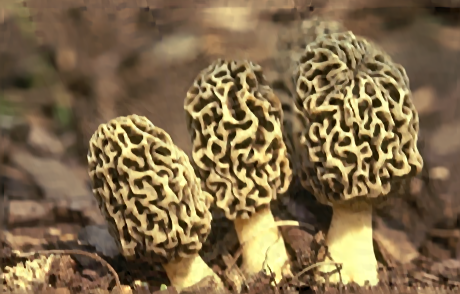}\llap{\includegraphics[height=0.03\textheight]{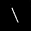}}};
		\spy[blue] on (-0.8,-0.5) in node[below,ultra thick] at (0,-1.5);
		\end{tikzpicture}
	}\\
	\vspace{-3mm}
	\setcounter{subfigure}{0}
	\subfloat[Groundtruth]{%
		\begin{tikzpicture}[spy using outlines={rectangle,magnification=8,width=4.2cm,height=2.5cm}]
		\node {\includegraphics[height=0.11\textheight]{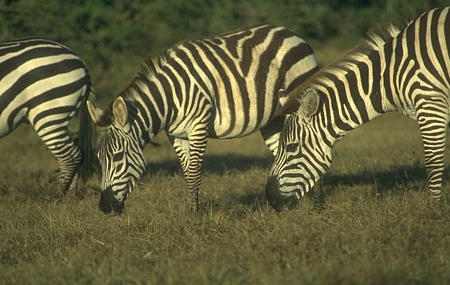}\llap{\includegraphics[height=0.03\textheight]{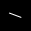}}};
		\spy[blue] on (1.5,0.7) in node[below,ultra thick] at (0,-1.5);
		\end{tikzpicture}
	}\hspace{-3mm}
	\subfloat[Perrone {\it et al.}~\cite{perrone_clearer_2016}]{%
		\begin{tikzpicture}[spy using outlines={rectangle,magnification=8,width=4.2cm,height=2.5cm}]
		\node {\includegraphics[height=0.11\textheight]{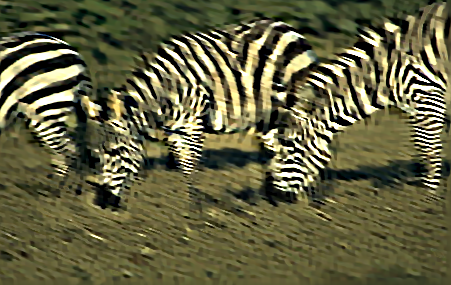}\llap{\includegraphics[height=0.03\textheight]{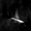}}};
		\spy[blue] on (1.5,0.7) in node[below,ultra thick] at (0,-1.5);
		\end{tikzpicture}
	}\hspace{-3mm}
	\subfloat[Nah {\it et al.}~\cite{nah_deep_2017}]{%
		\begin{tikzpicture}[spy using outlines={rectangle,magnification=8,width=4.2cm,height=2.5cm}]
		\node {\includegraphics[height=0.11\textheight]{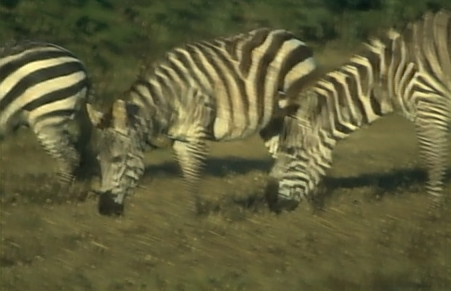}};
		\spy[blue] on (1.5,0.7) in node[below,ultra thick] at (0,-1.5);
		\end{tikzpicture}
	}\hspace{-3mm}
	\subfloat[DUBLID]{%
		\begin{tikzpicture}[spy using outlines={rectangle,magnification=8,width=4.2cm,height=2.5cm}]
		\node {\includegraphics[height=0.11\textheight]{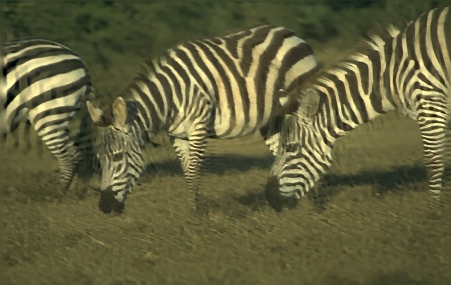}\llap{\includegraphics[height=0.03\textheight]{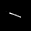}}};
		\spy[blue] on (1.5,0.7) in node[below,ultra thick] at (0,-1.5);
		\end{tikzpicture}
	}
	\caption{Qualitative comparisons on the BSDS500~\cite{martin_database_2001} dataset. The blur kernels are placed at the right bottom corner. DUBLID recovers the kernel at higher accuracy and therefore the estimated images are more faithful to the groundtruth.}\label{fig:compare_linear}\vspace{-15pt}
\end{figure*}

Fig.~\ref{fig:compare_linear} shows four example images and kernels for a qualitative
comparison. The two top-performing methods, Perrone {\it et al.}~\cite{perrone_clearer_2016} and
Nah {\it et al.}~\cite{nah_deep_2017}, are also included as representatives of iterative
methods and deep learning methods, respectively. Although~\cite{perrone_clearer_2016} can roughly infer the directions of the blur kernels, the
recovered coefficients clearly differ from the groundtruth as evidenced by the
spread-out branches.  Consequently, destroyed local structures and false colors
are clearly observed in the reconstructed images. Nah {\it et al.}'s method~\cite{nah_deep_2017}
does not suffer from false colors, yet the recovered images appear blurry. In
contrast, DUBLID recovers kernels close to the groundtruth, and produces
significantly fewer visually objectionable artifacts in the recovered images.\vspace{-10pt}

\begin{table*}
	\centering
	\caption{Quantitative comparison over an average of 200 images and 4 kernels. The best scores are in bold fonts.}\label{tab:linear}
	\begin{tabular}{ccccccc}
		\toprule
		Metrics & DUBLID  & Perrone {\it et al.}~\cite{perrone_clearer_2016} & Nah {\it et al.}~\cite{nah_deep_2017} &  Xu {\it et al.}~\cite{xu_motion_2018} & Kupyn \textit{et al.}~\cite{kupyn_deblurgan_2018}\\
		\midrule
		PSNR (dB) & $\mathbf{27.30}$ & $22.23$ & $24.82$ & $24.02$ & $23.98$\\
		\midrule
		ISNR (dB) & $\mathbf{4.45}$ & $2.06$ & $1.92$ & $1.12$ & $1.05$\\
		\midrule
		SSIM & $\mathbf{0.88}$ & $0.76$ & $0.80$ & $0.78$ & $0.78$\\
		\midrule
		RMSE $\left(\times 10^{-3}\right)$ & $\mathbf{1.67}$ & $5.21$ & $-$ & $2.40$ & $-$\\
		\bottomrule
	\end{tabular}\vspace{-15pt}
\end{table*}

\begin{table*}
\centering
\caption{Quantitative comparison over an average of 4 images and 8 kernels from~\cite{levin_understanding_2011}.}\label{tab:levin}
\begin{tabular}{ccccccc}
	\toprule
	& DUBLID & Perrone {\it et al.}~\cite{perrone_clearer_2016} & Nah {\it et al.}~\cite{nah_deep_2017} & Chakrabarti {\it et al.}~\cite{chakrabarti_neural_2016} & Xu {\it et al.}~\cite{xu_motion_2018} & Kupyn \textit{et al.}~\cite{kupyn_deblurgan_2018}\\
	\midrule
	PSNR (dB) & $\mathbf{27.15}$ & $26.79$ & $24.51$ & $23.21$ & $26.75$ & $23.98$\\
	\midrule
	ISNR (dB) & $\mathbf{3.79}$ & $3.63$ & $1.35$ & $0.06$ & $3.59$ & $0.43$\\
	\midrule
	SSIM & $\mathbf{0.89}$ & $\mathbf{0.89}$ & $0.81$ & $0.81$ & $\mathbf{0.89}$ & $0.80$\\
	\midrule
	RMSE $\left(\times 10^{-3}\right)$ & $3.87$ & $\mathbf{3.83}$ & $-$ & $4.33$ & $3.98$ & $-$\\
	\bottomrule
\end{tabular}\vspace{-15pt}
\end{table*}

\begin{table*}
\centering
\caption{Quantitative comparison over an average of 80 images and 8 nonlinear motion kernels from~\cite{sun_edge-based_2013}.}\label{tab:sun}
\begin{tabular}{ccccccc}
	\toprule
	& DUBLID & Perrone {\it et al.}~\cite{perrone_clearer_2016} & Nah {\it et al.}~\cite{nah_deep_2017} & Chakrabarti {\it et al.}~\cite{chakrabarti_neural_2016} & Xu {\it et al.}~\cite{xu_motion_2018} & Kupyn \textit{et al.}~\cite{kupyn_deblurgan_2018}\\
	\midrule
	PSNR (dB) & $\mathbf{29.91}$ & $29.82$ & $26.98$ & $29.86$ & $26.55$ & $25.84$\\
	\midrule
	ISNR (dB) & $\mathbf{4.11}$ & $4.02$ & $0.86$ & $4.06$ & $0.43$ & $0.15$\\
	\midrule
	SSIM & $\mathbf{0.93}$ & $0.92$ & $0.85$ & $0.91$ & $0.87$ & $0.83$\\
	\midrule
	RMSE $\left(\times 10^{-3}\right)$ & $\mathbf{2.33}$ & $2.68$ & $-$ & $2.72$ & $2.79$ & $-$ \\
	\bottomrule
\end{tabular}\vspace{-5pt}
\end{table*}

\subsection{Evaluation on Non-linear Kernels}\label{subsec:nonlinear}
\begin{figure}
	\centering
	\subfloat{\includegraphics[width=0.11\linewidth]{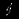}}\,\!
	\subfloat{\includegraphics[width=0.11\linewidth]{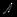}}\,\!
	\subfloat{\includegraphics[width=0.11\linewidth]{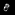}}\,\!
	\subfloat{\includegraphics[width=0.11\linewidth]{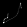}}\,\!
	\subfloat{\includegraphics[width=0.11\linewidth]{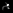}}\,\!
	\subfloat{\includegraphics[width=0.11\linewidth]{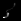}}\,\!
	\subfloat{\includegraphics[width=0.11\linewidth]{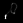}}\,\!
	\subfloat{\includegraphics[width=0.11\linewidth]{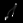}}\,\!\vspace{-3pt}
	\caption{Examples of realistic non-linear kernels~\cite{levin_understanding_2011}.\vspace{-10pt}}\label{fig:levin_kernels}
\end{figure}

It has been observed in several previous
works~\cite{levin_understanding_2011,kohler_recording_2012} that realistic
motion kernels often have non-linear shapes due to irregular camera motions,
such as those shown in Fig.~\ref{fig:levin_kernels}. Therefore, the capability
to handle such kernels is crucial for a blind motion deblurring method.

We generate training kernels by interpolating the paths provided
by~\cite{kohler_recording_2012} and those created by ourselves: specifically, we
record the camera motion trajectories using the Vicon system, and then
interpolate the trajectories spatially to create motion kernels. We further
augment these kernels by scaling over 4 different scales and rotating over 8
directions. In this way, we build around $30,000$ training kernels in total\footnote{To re-emphasize, all learning based methods use the same training-test configuration for fairness in comparison.}.
The blurred images for training are synthesized by randomly picking a kernel
and convolving with it. Gaussian noise of standard deviation $0.01$ is again
added. We use the standard image set from~\cite{levin_understanding_2011}
(comprising 4 images and 8 kernels) and from~\cite{sun_edge-based_2013}
(comprising 80 images and 8 kernels) as the test sets. The average scores for
both datasets are presented in Table~\ref{tab:levin} and Table~\ref{tab:sun},
respectively. In both datasets, DUBLID emerges overall as the best method. The method of Chakrabarti {\it et al.}~\cite{chakrabarti_neural_2016} performs second best in Table ~\ref{tab:sun}. In Table~\ref{tab:levin}, 
Perrone {\it et al.}~\cite{perrone_clearer_2016}  and the recent deep learning method of Xu {\it et al.} ~\cite{xu_motion_2018} perform comparably and mildly worse than DUBLID. DUBLID however achieves the deblurring at a significantly lower computational cost as verified in Section \ref{subsec:computation}.

\begin{figure*}[t]
	\centering
	\setcounter{subfigure}{0}
	\subfloat[Groundtruth]{%
		\begin{tikzpicture}[spy using outlines={rectangle,magnification=8,width=2.5cm,height=2cm}]
			\node {\includegraphics[height=0.09\textheight]{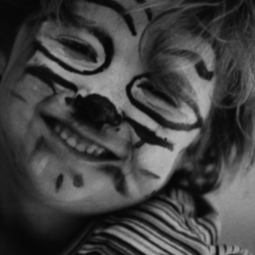}\llap{\includegraphics[height=0.03\textheight]{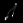}}};
			\spy[blue] on (0.3,0.2) in node[below,ultra thick] at (0,-1.4);
		\end{tikzpicture}
	}\hspace{-3mm}
	\subfloat[Perrone {\it et al.}~\cite{perrone_clearer_2016}]{%
		\begin{tikzpicture}[spy using outlines={rectangle,magnification=8,width=2.5cm,height=2cm}]
			\node {\includegraphics[height=0.09\textheight]{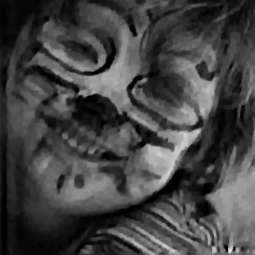}\llap{\includegraphics[height=0.03\textheight]{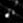}}};
			\spy[blue] on (0.3,0.2) in node[below,ultra thick] at (0,-1.4);
		\end{tikzpicture}
	}\hspace{-3mm}
	\subfloat[Nah {\it et al.}~\cite{nah_deep_2017}]{%
		\begin{tikzpicture}[spy using outlines={rectangle,magnification=8,width=2.5cm,height=2cm}]
			\node {\includegraphics[height=0.09\textheight]{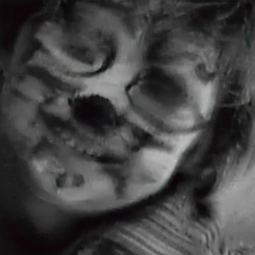}};
			\spy[blue] on (0.3,0.2) in node[below,ultra thick] at (0,-1.4);
		\end{tikzpicture}
	}\hspace{-3mm}
	\subfloat[Chakrabarti~\cite{chakrabarti_neural_2016}]{%
		\begin{tikzpicture}[spy using outlines={rectangle,magnification=8,width=2.5cm,height=2cm}]
			\node {\includegraphics[height=0.09\textheight]{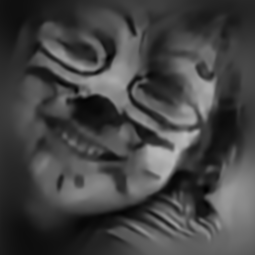}\llap{\includegraphics[height=0.03\textheight]{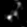}}};
			\spy[blue] on (0.3,0.2) in node[below,ultra thick] at (0,-1.4);
		\end{tikzpicture}
	}\hspace{-3mm}
	\subfloat[Xu {\it et al.}~\cite{xu_motion_2018}]{%
		\begin{tikzpicture}[spy using outlines={rectangle,magnification=8,width=2.5cm,height=2cm}]
			\node {\includegraphics[height=0.09\textheight]{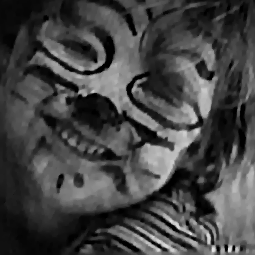}\llap{\includegraphics[height=0.03\textheight]{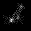}}};
			\spy[blue] on (0.3,0.2) in node[below,ultra thick] at (0,-1.4);
		\end{tikzpicture}
	}\hspace{-3mm}
	\subfloat[Kupyn {\it et al.}~\cite{kupyn_deblurgan_2018}]{%
		\begin{tikzpicture}[spy using outlines={rectangle,magnification=8,width=2.5cm,height=2cm}]
			\node {\includegraphics[height=0.09\textheight]{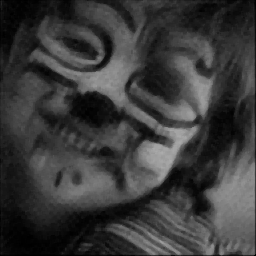}};
			\spy[blue] on (0.3,0.2) in node[below,ultra thick] at (0,-1.4);
		\end{tikzpicture}
	}\hspace{-3mm}
	\subfloat[DUBLID]{%
		\begin{tikzpicture}[spy using outlines={rectangle,magnification=8,width=2.5cm,height=2cm}]
			\node {\includegraphics[height=0.09\textheight]{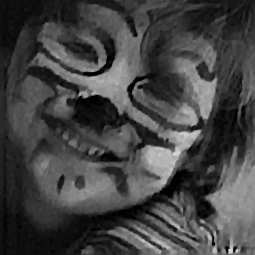}\llap{\includegraphics[height=0.03\textheight]{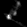}}};
			\spy[blue] on (0.3,0.2) in node[below,ultra thick] at (0,-1.4);
		\end{tikzpicture}
	}
	\caption{Qualitative comparisons on the dataset from~\cite{levin_understanding_2011}. The blur kernels are placed at the right bottom corner. DUBLID generates fewer artifacts and preserves more details than competing state of the art methods.}\label{fig:levin}\vspace{-18pt}
\end{figure*}

\begin{figure*}[t]
	\centering
	\setcounter{subfigure}{0}
	\subfloat[Groundtruth]{%
		\begin{tikzpicture}[spy using outlines={rectangle,magnification=8,width=2.2cm,height=2cm}]
			\node {\includegraphics[height=0.14\textheight]{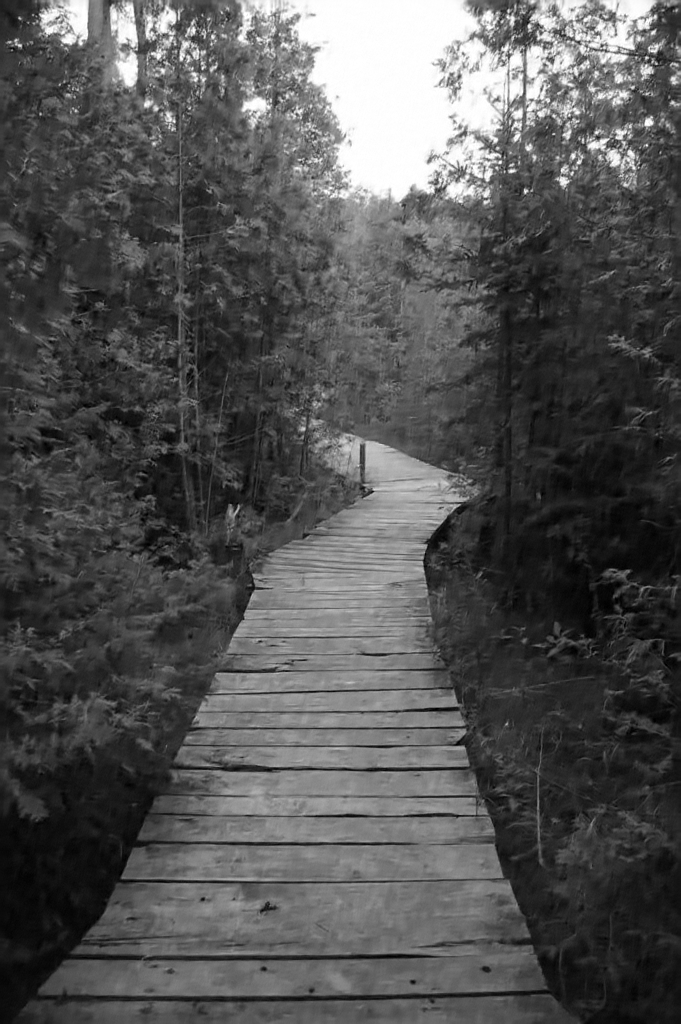}\llap{\includegraphics[height=0.03\textheight]{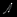}}};
			\spy[blue] on (0.3,0.2) in node[below,ultra thick] at (0,-1.9);
		\end{tikzpicture}
	}
	\subfloat[Perrone {\it et al.}~\cite{perrone_clearer_2016}]{%
		\begin{tikzpicture}[spy using outlines={rectangle,magnification=8,width=2.2cm,height=2cm}]
			\node {\includegraphics[height=0.14\textheight]{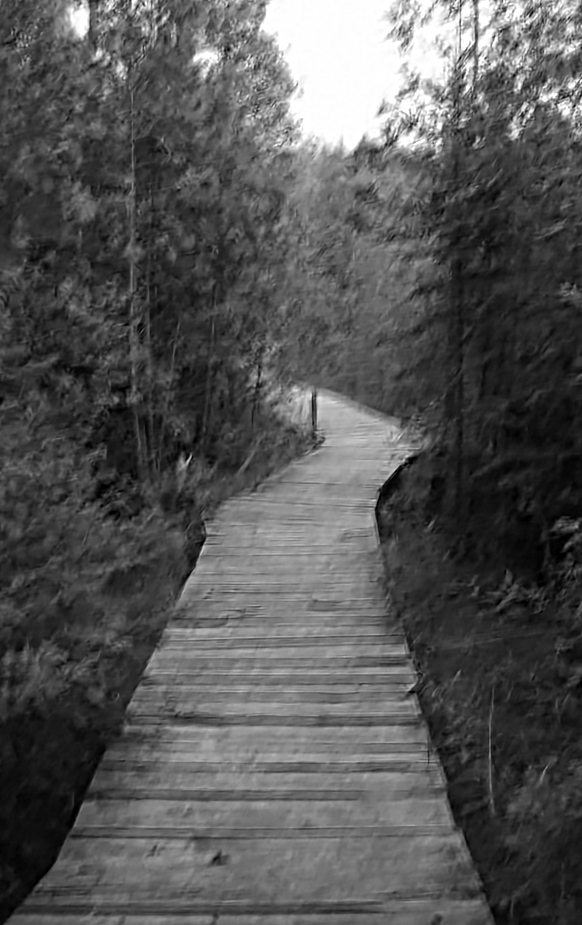}\llap{\includegraphics[height=0.03\textheight]{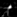}}};
			\spy[blue] on (0.3,0.2) in node[below,ultra thick] at (0,-1.9);
		\end{tikzpicture}
	}
	\subfloat[Nah {\it et al.}~\cite{nah_deep_2017}]{%
		\begin{tikzpicture}[spy using outlines={rectangle,magnification=8,width=2.2cm,height=2cm}]
			\node {\includegraphics[height=0.14\textheight]{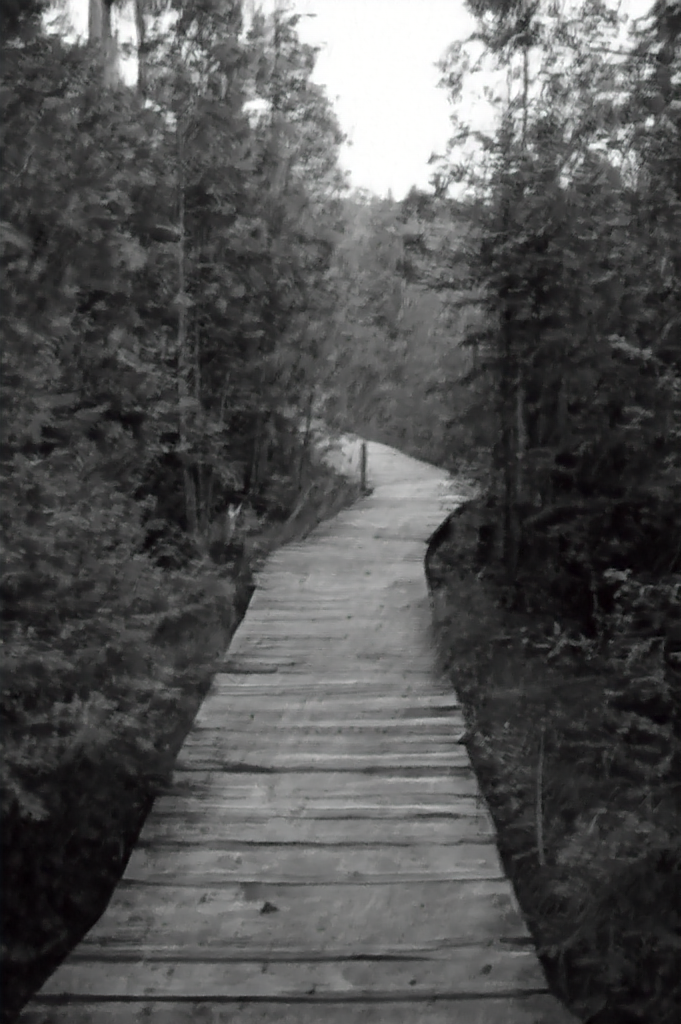}};
			\spy[blue] on (0.3,0.2) in node[below,ultra thick] at (0,-1.9);
		\end{tikzpicture}
	}
	\subfloat[Chakrabarti~\cite{chakrabarti_neural_2016}]{%
		\begin{tikzpicture}[spy using outlines={rectangle,magnification=8,width=2.2cm,height=2cm}]
			\node {\includegraphics[height=0.14\textheight]{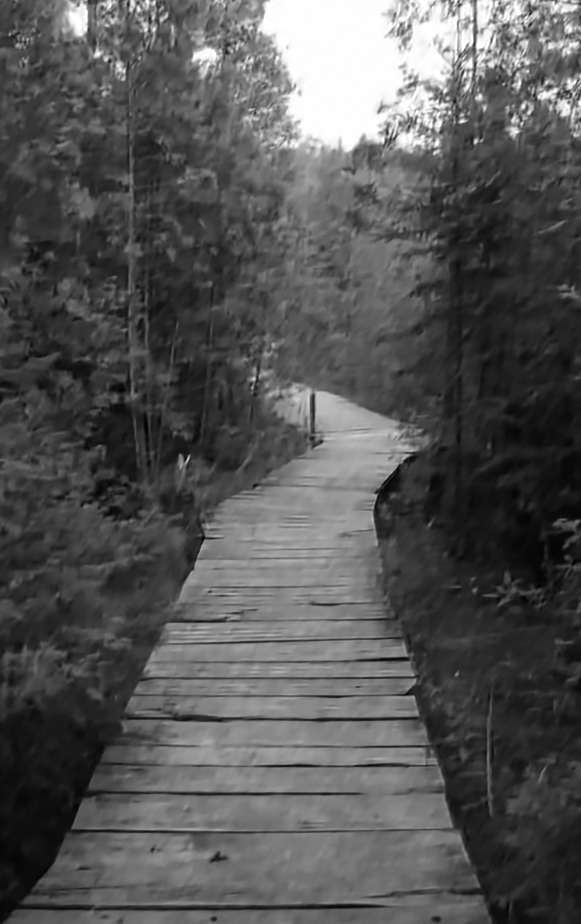}\llap{\includegraphics[height=0.03\textheight]{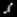}}};
			\spy[blue] on (0.3,0.2) in node[below,ultra thick] at (0,-1.9);
		\end{tikzpicture}
	}
	\subfloat[Xu {\it et al.}~\cite{xu_motion_2018}]{%
		\begin{tikzpicture}[spy using outlines={rectangle,magnification=8,width=2.2cm,height=2cm}]
			\node {\includegraphics[height=0.14\textheight]{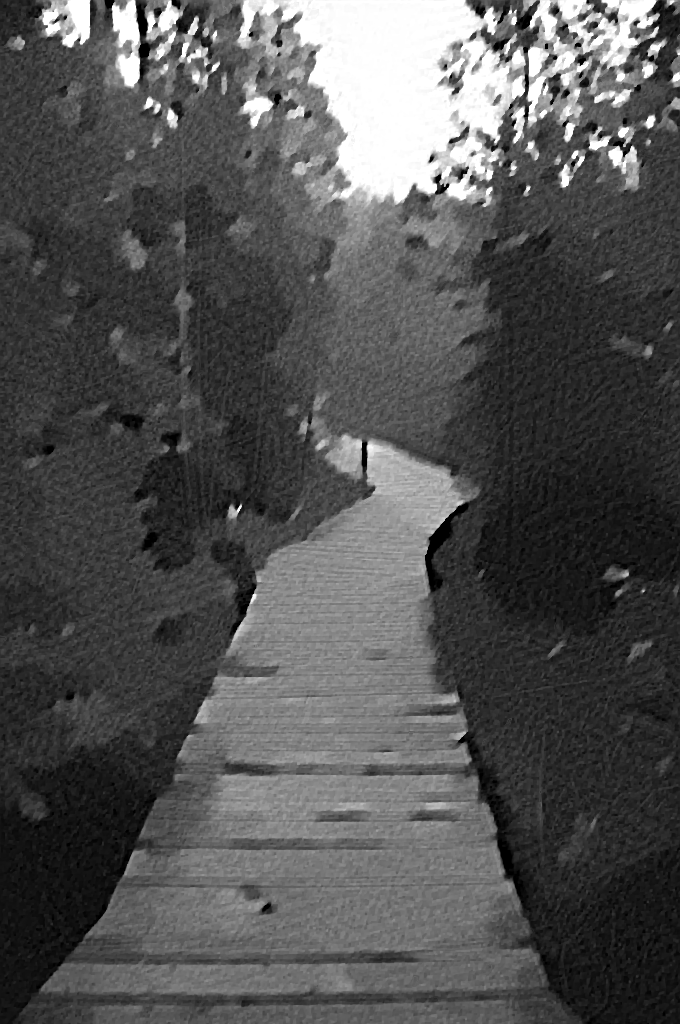}\llap{\includegraphics[height=0.03\textheight]{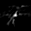}}};
			\spy[blue] on (0.3,0.2) in node[below,ultra thick] at (0,-1.9);
		\end{tikzpicture}
	}
	\subfloat[Kupyn {\it et al.}~\cite{kupyn_deblurgan_2018}]{%
		\begin{tikzpicture}[spy using outlines={rectangle,magnification=8,width=2.2cm,height=2cm}]
			\node {\includegraphics[height=0.14\textheight]{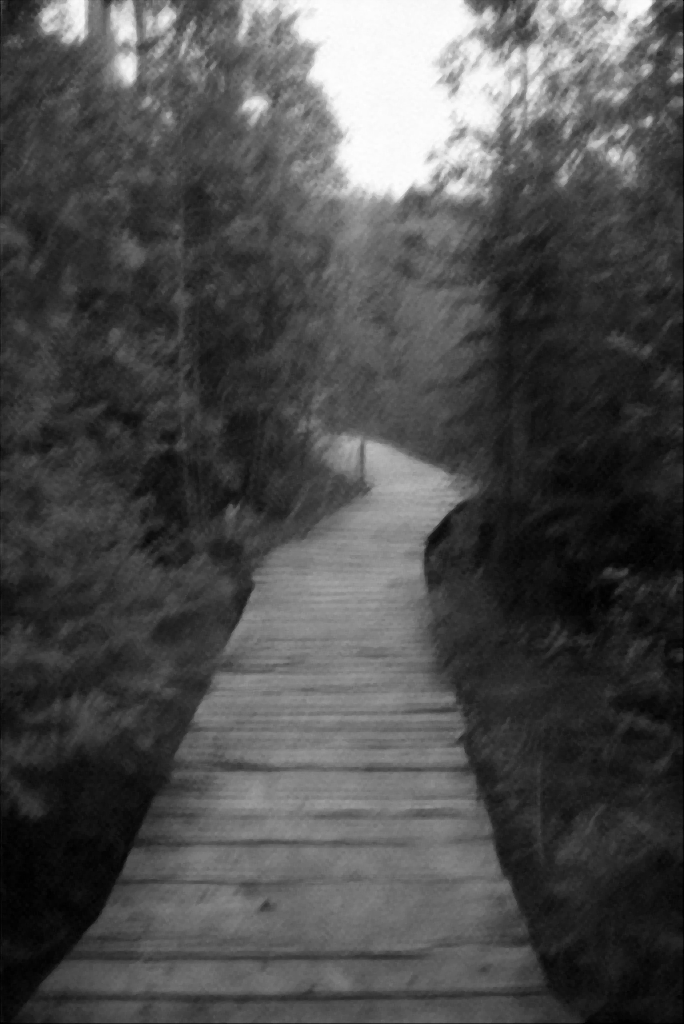}};
			\spy[blue] on (0.3,0.2) in node[below,ultra thick] at (0,-1.9);
		\end{tikzpicture}
	}
	\subfloat[DUBLID]{%
		\begin{tikzpicture}[spy using outlines={rectangle,magnification=8,width=2.2cm,height=2cm}]
			\node {\includegraphics[height=0.14\textheight]{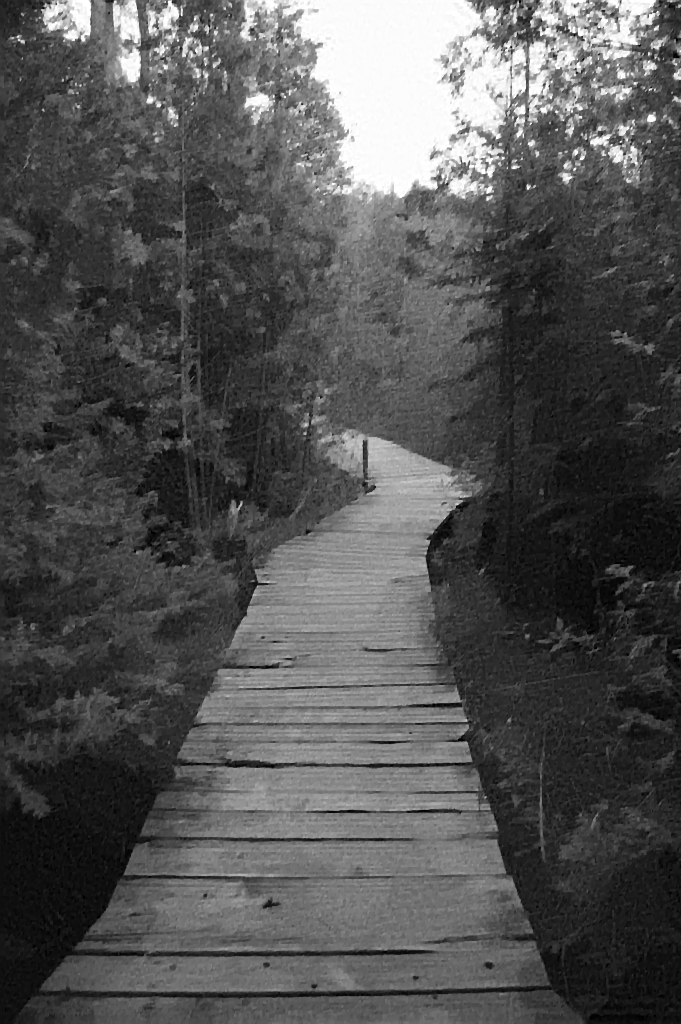}\llap{\includegraphics[height=0.03\textheight]{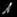}}};
			\spy[blue] on (0.3,0.2) in node[below,ultra thick] at (0,-1.9);
		\end{tikzpicture}
	}
	\caption{Qualitative comparisons on the dataset from~\cite{sun_edge-based_2013}. The blur kernels are placed at the right bottom corner.}\label{fig:sun}\vspace{-15pt}
\end{figure*}

Visual examples are shown in Figs.~\ref{fig:levin}
and~\ref{fig:sun} for qualitative comparisons. It can be clearly
seen that DUBLID is capable of more faithfully recovering the kernels, and hence
produces reconstructed images of higher visual quality. In particular, DUBLID preserves
local details better as shown in the zoom boxes of Figs.~\ref{fig:levin}
and~\ref{fig:sun} while providing sharper images than Nah {\it et al.}~\cite{nah_deep_2017}, Chakrabarti {\it et al.}~\cite{chakrabarti_neural_2016} and Kupyn {\it
et al.}~\cite{kupyn_deblurgan_2018}. Finally, DUBLID is free of visually objectionable artifacts
observed in Perrone {\it et al.}~\cite{perrone_clearer_2016} and Xu
{\it et al.}~\cite{xu_motion_2018}.\vspace{-10pt}

\subsection{Computational Comparisons Against State of the Art}\label{subsec:computation}

Table~\ref{tab:runtime} summarizes the execution (inference) times of each
method for processing a typical blurred image of resolution $480 \times 320$
and a blur kernel of size $31 \times 31$. The number of parameters for DUBLID
is estimated as follows: for $3\times 3$ filters $\bw_{ij}$, there are a total
of $L = 10$ layers and in each layer there are $C^2 = 16 \times 16$ filters,
which contribute to $3\times 3\times 16\times 16\times 10 \approx 2.3\times
10^4$ parameters. Other parameters have negligible dimensions compared with
$\bw_{ij}$ and thus do not contribute significantly.

We include measurements of running time on both CPU and GPU.\@ The $-$ symbol
indicates inapplicability. For instance, Chakrabarti {\em et
al.}~\cite{chakrabarti_neural_2016}  and Nah {\em et al.}~\cite{nah_deep_2017}
only provide GPU implementations of their work and likewise Perrone {\em et
al}'s iterative method~\cite{perrone_clearer_2016} is only implemented on a
CPU. Specifically, the two benchmark platforms are: 1.) Intel Core i7--6900K, 3.20GHz
CPU, 8GB of RAM, and 2.) an NVIDIA TITAN X GPU.\@ The results in
Table~\ref{tab:runtime} deliver two messages. First, the deep/neural network
based methods are faster than their iterative algorithm counterparts, which is
to be expected. Second, amongst the deep neural net methods DUBLID runs
significantly faster than the others on both GPU and CPU, largely  because it
has significantly fewer parameters as seen in the final row of
Table~\ref{tab:runtime}. Note that the number of parameters for competing deep
learning methods are computed based on the description in their respective
papers.


\begin{table*}
	\centering
	\caption{Running time comparisons over different methods. The image size is $480\times 320$ and the kernel size is $31\times 31$.\vspace{-3pt}}\label{tab:runtime}
	\begin{tabular}{cccccccc}
		\toprule
		& DUBLID  & Chakrabarti {\it et al.}~\cite{chakrabarti_neural_2016} & Nah {\it et al.}~\cite{nah_deep_2017} & Perrone {\it et al.}~\cite{perrone_clearer_2016} & Xu {\it et al.}~\cite{xu_motion_2018}  & Kupyn \textit{et al.}~\cite{kupyn_deblurgan_2018}\\
		\midrule
		CPU Time (s) & $\mathbf{1.47}$ & $-$ & $-$ & $1462.90$ & $6.89$ & $10.29$\\
		\midrule
		GPU Time (s) & $\mathbf{0.05}$ & $227.80$ & $7.32$ & $-$ & $2.01$ & $0.13$\\
		\midrule
		Number of Parameters & $\mathbf{2.3\times 10^4}$ & $1.1\times 10^8$ & $2.3\times 10^7$ & $-$ & $6.0\times 10^6$ & $1.2\times 10^7$\\
		\bottomrule
	\end{tabular}\vspace{-10pt}
\end{table*}

\section{Conclusion}\label{sec:conclusion}
We propose an Algorithm Unrolling approach for Deep Blind image Deblurring (DUBLID). Our approach is based on recasting a generalized TV-regularized algorithm
into a neural network, and optimizing its parameters via a custom designed backpropogation procedure. Unlike
most existing neural network approaches, our technique has the benefit of
interpretability, while sharing the performance benefits of modern neural
network approaches. While some existing approaches excel for the case of linear kernels and others for non-linear, our method is versatile across a variety of scenarios and kernel choices -- as is verified both visually and quantitatively. Further, DUBLID requires much fewer parameters leading to significant computational benefits over iterative methods as well as competing deep learning techniques.\vspace{-5pt}

\appendices

\section{Gradients Computation by Back-Propagation}\label{sec:backpropagation}
Here we develop the back-propagation rules for computing the
gradients of DUBLID.\@ We will use $\bF$ to denote the DFT operator and
$\bF^\ast$ its adjoint operator, and $\bone$ is a
vector whose entries are all ones. $\bI$ refers to the identiy matrix. The symbols $\cI_{\{\}}$ means indicator vectors and
$\diag(\cdot)$ embeds the vector into a diagonal matrix. The operators $\bP_\bg$ and
$\bP_\bk$ are projections that restrict the operand into the domain of
the image and the kernel, respectively. Let $\cL$ be the cost function defined in~\eqref{eqn:loss}.
We derive its gradients w.r.t.\ its variables using the chain rule as follows:\vspace{-5pt}
\begin{align*}
	\nabla_{\bw_i^l}\mathcal{L}&=\nabla_{\bw_i^l}\by_i^l\nabla_{\by_i^l}\cL=\bR_{\bw_i^l}\bF\diag\left(\widehat{\by_i^{l+1}}\right)\bF^\ast\nabla_{\by_i^l}\mathcal{L},\\
	\nabla_{\zeta_i^l}\mathcal{L}&=\nabla_{\zeta_i^l}\bz_i^{l+1}\nabla_{\bz_i^{l+1}}\cL\\
								 &\resizebox{.9\linewidth}{!}{$=\left[\frac{\widehat{\bk^l}^\ast\odot\left(\widehat{\bk^l}\odot\widehat{\bg_i^l}-\widehat{\by_i^l}\right)}{\left(\left|\widehat{\bk^l}\right|^2+\zeta_i^l\right)^2}\right]^T\bF^\ast\left(\cI_{\left\{\left|\bP_\bg\bg_i^{l+1}\right|>b_i^l\right\}}\odot\nabla_{\bz_i^{l+1}}\mathcal{L}\right)$},
	\end{align*}\vspace{-6pt}
	\begin{align*}
	\resizebox{.99\linewidth}{!}{$\nabla_{b_i^l}\mathcal{L}=\nabla_{b_i^l}\bz_i^{l+1}\nabla_{\bz_i^{l+1}}\cL={\left(\cI_{\left\{\bg_i^{l+1}<-b_i^l\right\}}-\cI_{\left\{\bg_i^{l+1}>b_i^l\right\}}\right)}^T\nabla_{\bz_i^{l+1}}\mathcal{L}$},
\end{align*}
where $\bR_{\bw_i^l}$ is the operator that extracts the components lying in the support of $\bw_i^l$. Again using the chain rule,\vspace{-5pt}
\begin{align}
\frac{\partial \mathcal{L}}{\partial\bk^l}&=\frac{\partial \mathcal{L}}{\partial\bz_i^{l+1}}\frac{\partial\bz_i^{l+1}}{\partial\bk^l},~~
\frac{\partial \mathcal{L}}{\partial\bz_i^l}=\frac{\partial \mathcal{L}}{\bz_i^{l+1}}\frac{\partial\bz_i^{l+1}}{\partial\bz_i^l}+\frac{\partial \mathcal{L}}{\partial\bk^l}\frac{\partial\bk^l}{\partial\bz_i^l},\nonumber\\
\frac{\partial \mathcal{L}}{\partial\by_i^l}&=\frac{\partial \mathcal{L}}{\bz_i^{l+1}}\frac{\partial\bz_i^{l+1}}{\partial\by_i^l}+\frac{\partial \mathcal{L}}{\partial\bk^{l+1}}\frac{\partial\bk^{l+1}}{\partial\by_i^l}+\frac{\partial \mathcal{L}}{\partial\by_i^{l-1}}\frac{\partial\by_i^{l-1}}{\partial\by_i^l}.\label{eq:summary}
\end{align}
We next derive each individual term in~\eqref{eq:summary} as follows:
\begin{align}
	\frac{\partial\bz_i^{l+1}}{\partial\widehat{\bg_i^{l+1}}}&=\frac{\partial\bz_i^{l+1}}{\bg_i^{l+1}}\frac{\partial\bg_i^{l+1}}{\partial\widehat{\bg_i^{l+1}}}=\diag\left(\cI_{\left\{\left|\bP_\bg\bg_i^{l+1}\right|>b_i^l\right\}}\right)\bF^\ast,\nonumber\\
	\frac{\partial\bz_i^{l+1}}{\partial\bz_i^l}&=\frac{\partial\bz_i^{l+1}}{\partial\widehat{\bg_i^{l+1}}}\frac{\partial\widehat{\bg_i^{l+1}}}{\partial\widehat{\bz_i^l}}\frac{\partial\widehat{\bz_i^l}}{\partial\bz_i^l}\label{eq:dzdg}\\
											   &=\diag\left(\cI_{\left\{\left|\bP_\bg\bg_i^{l+1}\right|>b_i^l\right\}}\right)\bF^\ast\diag\left(\frac{\zeta_i^l}{\left|\widehat{\bk^l}\right|^2+\zeta_i^l}\right)\bF,\nonumber\\
	\frac{\partial\bz_i^{l+1}}{\partial\by_i^l}&=\frac{\partial\bz_i^{l+1}}{\partial\widehat{\bg_i^{l+1}}}\frac{\partial\widehat{\bg_i^{l+1}}}{\partial\widehat{\by_i^l}}\frac{\partial\widehat{\by_i^l}}{\partial\by_i^l}\label{eq:dzdy}\\
											   &=\diag\left(\cI_{\left\{\left|\bP_\bg\bg_i^{l+1}\right|>b_i^l\right\}}\right)\bF^\ast\diag\left(\frac{\widehat{\bk^l}^\ast}{\left|\widehat{\bk^l}\right|^2+\zeta_i^l}\right)\bF\nonumber,
\end{align}	\vspace{-8pt}	
	  
and
\begin{align}
	\frac{\partial\bz_i^{l+1}}{\partial\bk^l}&=\frac{\partial\bz_i^{l+1}}{\partial\widehat{\bg_i^{l+1}}}\left(\frac{\partial\widehat{\bg_i^{l+1}}}{\partial\widehat{\bk^l}}\frac{\partial\widehat{\bk_i^l}}{\partial\bk_i^l}+\frac{\partial\widehat{\bg_i^{l+1}}}{\partial\widehat{\bk^l}^\ast}\frac{\partial\widehat{\bk_i^l}^\ast}{\partial\bk_i^l}\right)\label{eq:dzdk}\\
											 &=\diag\left(\cI_{\left\{\left|\bP_\bg\bg_i^{l+1}\right|>b_i^l\right\}}\right)\bF^\ast\nonumber\\
											 &\resizebox{.9\linewidth}{!}{$\left[\diag\left(\frac{\zeta_i^l\widehat{\by_i^l}}{\left(\left|\widehat{\bk^l}\right|^2+\zeta_i^l\right)^2}\right)\bF^\ast-\diag\left(\frac{\left(\widehat{\bk^l}^\ast\right)^2\odot\widehat{\by_i^l}}{\left(\left|\widehat{\bk^l}\right|^2+\zeta_i^l\right)^2}\right)\bF\right]$},\nonumber
\end{align}\vspace{-5pt}
\begin{align*}
	\frac{\partial\bk^{l+1}}{\partial\widehat{\bk^{l+\frac{1}{3}}}}&=\frac{\partial\bk^{l+1}}{\partial\bk^{l+\frac{2}{3}}}\frac{\partial\bk^{l+\frac{2}{3}}}{\partial\bk^{l+\frac{1}{3}}}\frac{\partial\bk^{l+\frac{1}{3}}}{\partial\widehat{\bk^{l+\frac{1}{3}}}}\\
																   &=\frac{\bI\left(\bone^T\bk^{l+\frac{2}{3}}\right)-\bk^{l+\frac{2}{3}}\bone^T}{{\left(\bone^T\bk^{l+\frac{2}{3}}\right)}^2}\diag\left(\cI_{\left\{\bP_\bk\bk^{l+\frac{1}{3}}>0\right\}}\right)\bF^\ast,
\end{align*}	\vspace{-10pt}		  
\begin{align}
	\frac{\partial\bk^{l+1}}{\partial\by_i^l}&=\frac{\partial\bk^{l+1}}{\partial\widehat{\bk^{l+\frac{1}{3}}}}\frac{\widehat{\bk^{l+\frac{1}{3}}}}{\widehat{\by_i^l}}\frac{\widehat{\by_i^l}}{\by_i^l}=\frac{\bI\left(\bone^T\bk^{l+\frac{2}{3}}\right)-\bk^{l+\frac{2}{3}}\bone^T}{{\left(\bone^T\bk^{l+\frac{2}{3}}\right)}^2}\cdot\label{eq:dkdy}\\
	&\diag\left(\cI_{\left\{\bP_\bk\bk^{l+\frac{1}{3}}>0\right\}}\right)\bF^\ast\diag\left(\frac{\sum_{i=1}^C\widehat{\bz_i^{l+1}}^\ast}{\sum_{i=1}^C\left|\bz_i^{l+1}\right|^2+\varepsilon}\right)\bF,\nonumber
\end{align}	\vspace{-10pt}	  
\begin{align}
	\frac{\partial\bk^{l+1}}{\partial\bz_i^{l+1}}&=\frac{\partial\bk^{l+1}}{\partial\widehat{\bk^{l+\frac{1}{3}}}}\left(\frac{\partial\widehat{\bk^{l+\frac{1}{3}}}}{\partial\widehat{\bz_i^{l+1}}}\frac{\partial\widehat{\bz_i^{l+1}}}{\partial\bz_i^{l+1}}+\frac{\partial\widehat{\bk^{l+\frac{1}{3}}}}{\partial\widehat{\partial\bz_i^{l+1}}^\ast}\frac{\widehat{\partial\bz_i^{l+1}}^\ast}{\partial\bz_i^{l+1}}\right)\label{eq:dkdz}\\
												 &=\frac{\bI\left(\bone^T\bk^{l+\frac{2}{3}}\right)-\bk^{l+\frac{2}{3}}\bone^T}{\left(\bone^T\bk^{l+\frac{2}{3}}\right)^2}\diag\left(\cI_{\left\{\bP_\bk\bk^{l+\frac{1}{3}}>0\right\}}\right)\bF^\ast\cdot\nonumber
\end{align}		\vspace{-10pt}	  
\begin{align*}
												 &\left[-\diag\left(\frac{\left(\sum_{j=1}^C\widehat{\bz_j^{l+1}}^\ast\odot\widehat{\by_j^l}\right)\odot\widehat{\bz_i^{l+1}}^\ast}{\left(\sum_{j=1}^C\left|\widehat{\bz_j^{l+1}}\right|^2+\varepsilon\right)^2}\right)\bF\right.\\
												 &\resizebox{.9\linewidth}{!}{$+\left.\diag\left(\frac{\widehat{\by_i^l}\odot\left(\sum_{j=1}^C\left|\widehat{\bz_j^{l+1}}\right|^2+\varepsilon\right)-\left(\sum_{j=1}^C\widehat{\bz_j^{l+1}}^\ast\odot\widehat{\by_j^l}\right)\odot\widehat{\bz_i^{l+1}}}{\left(\sum_{j=1}^C\left|\widehat{\bz_j^{l+1}}\right|^2+\varepsilon\right)^2}\right)\bF^\ast\right]$},
 \end{align*}			\vspace{-5pt}  
 \begin{align}
	 \frac{\partial\by_i^{l-1}}{\partial\by_i^l}&=\frac{\partial\by_i^{l-1}}{\partial\widehat{\by_i^{l-1}}}\frac{\partial\widehat{\by_i^{l-1}}}{\partial\widehat{\by_i^l}}\frac{\partial\widehat{\by_i^l}}{\partial\by_i^l}=\bF^\ast\diag\left(\widehat{\bw_i^{l-1}}\right)\bF.\label{eq:dydy}
\end{align}			\vspace{-5pt}  

Plugging~\eqref{eq:dzdg}~\eqref{eq:dzdy}~\eqref{eq:dzdk}~\eqref{eq:dkdy}~\eqref{eq:dkdz}~\eqref{eq:dydy} into~\eqref{eq:summary}, we obtain
\begin{align*}
	\nabla_{\bk^l}\mathcal{L}&\resizebox{.9\linewidth}{!}{$=\left[\bF^\ast\diag\left(\frac{\zeta_i^l\widehat{\by_i^l}}{\left(\left|\widehat{\bk^l}\right|^2+\zeta_i^l\right)^2}\right)-\bF\diag\left(\frac{\left(\widehat{\bk^l}^\ast\right)^2\odot\widehat{\by_i^l}}{\left(\left|\widehat{\bk^l}\right|^2+\zeta_i^l\right)^2}\right)\right]$}\\
	&\bF^\ast\left(\cI_{\left\{\left|\bP_\bg\bg_i^{l+1}\right|>b_i^l\right\}}\odot\nabla_{\bz_i^{l+1}}\mathcal{L}\right)
\end{align*}\vspace{-10pt}

\begin{align*}
	\nabla_{\bg_i^l}\mathcal{L}&=\bF\diag\left(\frac{\zeta_i^l}{\left|\widehat{\bk^l}\right|^2+\zeta_i^l}\right)\bF^\ast\left(\cI_{\left\{\left|\bP_\bg\bg_i^{l+1}\right|>b_i^l\right\}}\odot\nabla_{\bz_i^{l+1}}\mathcal{L}\right)\\
					 &\resizebox{.7\linewidth}{!}{$+\left[-\bF\diag\left(\frac{\left(\sum_{j=1}^\mathcal{L}\widehat{\bg_j^l}^\ast\odot\widehat{\by_j^{l-1}}\right)\odot\widehat{\bg_i^l}^\ast}{\left(\sum_{j=1}^\mathcal{L}\left|\widehat{\bg_j^l}\right|^2+\varepsilon\right)^2}\right)\right.$}\\
					 &\resizebox{.9\linewidth}{!}{$+\left.\bF^\ast\diag\left(\frac{\widehat{\by_i^{l-1}}\odot\left(\sum_{j=1}^\mathcal{L}\left|\widehat{\bg_j^l}\right|^2+\varepsilon\right)-\left(\sum_{j=1}^\mathcal{L}\widehat{\bg_j^l}^\ast\odot\widehat{\by_j^{l-1}}\right)\odot\widehat{\bg_i^l}}{\left(\sum_{j=1}^\mathcal{L}\left|\widehat{\bg_j^l}\right|^2+\varepsilon\right)^2}\right)\right]\bF^\ast$}		\\
					 &\resizebox{.9\linewidth}{!}{$\left(\frac{1}{\bone^T\bk^{l-\frac{1}{3}}}\cI_{\left\{\bP_\bk\bk^{l-\frac{2}{3}}>0\right\}}\odot\nabla_{\bk^l}C-\frac{\cI_{\left\{\bP_\bk\bk^{l-\frac{2}{3}}>0\right\}}{\bk^{l-\frac{1}{3}}}^T}{\left(\bone^T\bk^{l-\frac{1}{3}}\right)^2}\nabla_{\bk^l}\mathcal{L}\right)$}
\end{align*}\vspace{-20pt}

\begin{align*}
	\nabla_{\by_i^l}\mathcal{L}&\resizebox{.9\linewidth}{!}{$=\bF\diag\left(\frac{\widehat{\bk^l}^\ast}{\left|\widehat{\bk^l}\right|^2+\zeta_i^l}\right)\bF^\ast\left(\cI_{\left\{\left|\bP_\bg\bg_i^{l+1}\right|>b_i^l\right\}}\odot\nabla_{\bz_i^{l+1}}\mathcal{L}\right)$}\\
					 &+\bF\diag\left(\frac{\sum_{i=1}^\mathcal{L}\widehat{\bz_i^{l+1}}^\ast}{\sum_{i=1}^\mathcal{L}\left|\bz_i^{l+1}\right|^2+\varepsilon}\right)\bF^\ast	\\
					 &\resizebox{.9\linewidth}{!}{$\left(\frac{1}{\bone^T\bk^{l+\frac{2}{3}}}\cI_{\left\{\bP_\bk\bk^{l+\frac{1}{3}}>0\right\}}\odot\nabla_{\bk^{l+1}}C-\frac{\cI_{\left\{\bP_\bk\bk^{l+\frac{1}{3}}>0\right\}}{\bk^{l+\frac{2}{3}}}^T}{\left(\bone^T\bk^{l+\frac{2}{3}}\right)^2}\nabla_{\bk^{l+1}}C\right)$}		\\
					 &+\bF\diag\left(\widehat{\bw_i^{l-1}}\right)\bF^\ast\nabla_{\by_i^{l-1}}\mathcal{L}
\end{align*}\vspace{-10pt}


\bibliographystyle{IEEEbib}
\bibliography{Deconvolution,DeepLearning}

\end{document}